\documentclass{aastex62}

\newcommand{\Eexc}{$E_{\rm exc}$}
\newcommand{\Teff}{$T_{\rm eff}$}  
\newcommand{\kms}{km\,s$^{-1}$}
\def\ione{\,{\sc i}}
\def\ii{\,{\sc ii}}
\def\iii{\,{\sc iii}}
\def\iv{\,{\sc iv}}
\def\v{\,{\sc v}}

\shorttitle{Neon abundances in B-type stars }
\shortauthors{Alexeeva et al.}

\begin{document}

\title{{\bf Neon Abundances of B-stars in the Solar Neighborhood}}

\author{Sofya Alexeeva}
\affiliation{Key Lab of Optical Astronomy, National Astronomical Observatories, Chinese Academy of Sciences, A20 Datun Road, Chaoyang, Beijing, 100102, China \\}
\email{alexeeva@nao.cas.cn}
\nocollaboration

\author{Tianxiang Chen}
\affiliation{Shandong Provincial Key Laboratory of Optical Astronomy and Solar Terrestrial Environment, Institute of Space Sciences, Shandong University, Weihai, 264209, China \\}
\nocollaboration

\author{Tatyana Ryabchikova}
\affiliation{Institute of Astronomy of the RAS, Pyatnitskaya 48, 119017, Moscow, Russia \\}
\nocollaboration

\author{Weibin Shi}
\affiliation{Shandong Provincial Key Laboratory of Optical Astronomy and Solar Terrestrial Environment, Institute of Space Sciences, Shandong University, Weihai, 264209, China \\}
\nocollaboration

\author{Kozo Sadakane}
\affiliation{Astronomical Institute, Osaka Kyoiku University, Asahigaoka, Kashiwara shi, Osaka 582 8582, Japan\\}
\nocollaboration

\author{Masayoshi Nishimura}
\affiliation{2-6, Nishiyama Maruo, Yawata shi, Kyoto 614 8353, Japan \\}
\nocollaboration

\author{Gang Zhao}
\affiliation{Key Lab of Optical Astronomy, National Astronomical Observatories, Chinese Academy of Sciences, A20 Datun Road, Chaoyang, Beijing, 100102, China \\}
\email{gzhao@nao.cas.cn}

\begin{abstract}

We constructed a comprehensive model atom for Ne\ione\ -- Ne\ii\ using the most up-to-date atomic data available and evaluated the non-local thermodynamic equilibrium (NLTE) line formation for Ne\ione\ and Ne\ii\ in classical 1D models representing the atmospheres of B-type stars.
   We find that the large NLTE strengthening of the Ne\ione\ lines corresponding to the 2p$^5$3p $-$ 2p$^5$3s transitions array occurs due to extremely small photoionization cross-sections of 2p$^5$3s levels that leads to strong overpopulation of these levels relative to their LTE populations. 
  The deviations from LTE for the most Ne\ii\ lines are small and do not exceed 0.11~dex in the absolute value. 
   We analysed 20 lines of Ne\ione\ and 13 lines of Ne\ii\ for twenty-four B-type stars in the temperature range of 10\,400 $\le$ \Teff\ $\le$ 33\,400~K. For five stars, the NLTE leads to consistent abundances of Ne\ione\ and Ne\ii, while the difference in LTE abundance can reach up to 0.50~dex.
   The using of the experimental oscillator strengths recently measured by Piracha et al. (2015) leads to smaller line-by-line scatter for the most investigated stars. The averaged neon abundance in twenty-four B-type stars in solar neighborhood is 8.02$\pm$0.05. 
This value may provide indirect constraints on solar photospheric neon abundance.

\end{abstract}


\section{Introduction} \label{sec:intro}

  The Sun is the closest star and its chemical composition serves as a zero point in the frame of abundance references for most astronomical objects. 
  The abundance of neon in the solar atmosphere is highly important for our understanding of the solar structure, however it is still not well constrained.
  Meteoritic studies cannot provide the actual neon abundance in the solar system, because neon is a noble gas and it is not retained in CI chondrite meteorites.
  The solar abundance of neon also cannot be directly estimated from the analysis of photospheric Ne\ione\ lines, which are absent in the solar spectrum due to their high excitation energies. 
  What is the neon abundance of the Sun? 
    The solar abundances of neon and oxygen have been a subject of debate during the last decade, after \citet{2005ASPC..336...25A} summarized the chemical composition of the Sun with improved photospheric models. Their revision of the solar abundances resulted in 0.2$-$0.3 dex decrease of CNO abundances and hence the overall solar metallicity, that put the predictions of the depth of the convection zone, helium abundance, density, and sound speed to the serious disagreement with helioseismological constraints \citep{2005ApJ...618.1049B, 2008PhR...457..217B}.   
   This problem was named as 'Solar Model Problem' by \citet{2005ApJ...618.1049B} and it still remains unsolved \citep{2017ApJ...839...55V}, although, asteroseismology has been established as a standard approach to study solar-type stars. 
  One of the ways to bring the solar model into agreement with the helioseismology is to increase the neon abundance. 
  Solar model calculations by \citet{2005ApJ...631.1281B} showed that log~$\epsilon_{\rm Ne}$ = 8.29$\pm$0.05 (or Ne/O = 0.42) would be enough for this purpose.
  However, more careful investigations of helioseismology models suggest that 
only an increase of the neon abundance cannot fully remove the conflict between helioseismic data and the predictions of solar interiors models \citep{2006ApJ...649..529D}. 
  
  The solar Ne/O abundance ratio is commonly estimated indirectly on the basis of X ray and UV spectroscopy of the solar corona, solar flares, and solar wind.
  However, solar plasma can be affected by such processes as first ionization potential (FIP) effect \citep{1973PhRvL..31..650H, 2000RvGeo..38..247B}, gravitational settling \citep{1970SoPh...12..458G},
  Coulomb drag associated with the outgoing proton flux \citep{2000JGR...105...47B} and ion-neutral separation. 
 All of these factors make interpretations
more complicated. 
  The model of fractionation is still needed to be constrained for obtaining the solar abundances from those measured in the corona and the solar wind \citep{2014ApJ...789...60S}.  
  For example, the FIP fractionation can lead to systematic differences between the composition of the corona and the photosphere, and there is 
  evidence for spatial and time variability in the composition of various coronal features \citep{2003SSRv..107..665F}.
  \citet{2015ApJ...800..110L} measured the absolute abundance in the corona for both oxygen and neon, and found that both elements are affected by the FIP effect.
  They concluded that the Ne/O ratio is not constant in the solar atmosphere, both in time and at different heights, and that it cannot be reliably used to infer the neon abundance in the photosphere.
  
  A large number of measurements of Ne/O ratio were presented in the literature, and the highest ratio value exceeds the lowest one by six times. 
  The lowest value, Ne/O = 0.07$\pm$0.01, was obtained from the measurements of the fast solar wind in polar coronal holes \citep{2007SSRv..130..139G}.
  The highest ratio Ne/O = 0.41, was derived by \citet{2005Natur.436..525D}, who measured Ne/O ratio in a sample of nearby solar-type stars from their X-ray spectra. 
  They suggested that the same ratio might also be appropriate for the Sun, and in this case the solar interior models could be brought back into agreement with the helioseismology measurements. 
  \citet{2008A&A...486..995R} found an evidence for a trend of higher Ne/O ratios with increasing stellar activity level in the sense that stars with the higher activity level show the higher Ne/O ratio. 
  Since the solar behavior appears to be rather typical for low activity stars, \citet{2008A&A...486..995R} suggested the Ne/O $\approx$ 0.2.
  According to \citet{2011ApJ...743...22D}, neon saturation may increase the underlying stellar photospheric compositions, so the low activity coronae, 
  including that of the Sun, are generally depleted in neon.  
  Another solar wind ratio value of Ne/O=0.14, lying between the highest and the lowest values, was derived by \citet{2007A&A...471..315B}, where the solar neon abundance
  was calculated as log~$\epsilon_{\rm Ne}$ = 7.96$\pm$0.13, if the oxygen abundance log~$\epsilon_{\rm O}$ = 8.87$\pm$0.11 is adopted. 
  
  \citet{2005A&A...439..361Y} demonstrated that the average quiet Sun does not show any FIP effect, and found a Ne/O ratio of 0.175$\pm$0.031 for the quiet Sun \citep{2005A&A...444L..45Y}. This ratio was adopted by \citet{2009ARA&A..47..481A} to convert their photospheric O abundance to a Ne content, and the value log~$\epsilon_{\rm Ne}$ = 7.93$\pm$0.10 was recommended. 
  Recently, \citet{2018ApJ...855...15Y} revised the previous analyses of \citet{2005A&A...439..361Y,2005A&A...444L..45Y} using updated atomic data and concluded that if the Ne/O ratio is interpreted as reflecting the photospheric ratio, then the photospheric neon abundance is 8.08$\pm$0.09 or 8.15$\pm$0.10, 
  according to whether the oxygen abundances of \citet{2009ARA&A..47..481A} or \citet{2011SoPh..268..255C} are used. 
  Absolute abundances of neon are generally calculated relative to a reference element such as oxygen, but it also can be obtained directly. 
  \citet{2007ApJ...659..743L} measured the absolute abundance of neon in the solar atmosphere, log~$\epsilon_{\rm Ne}$ = 8.11$\pm$0.12, from UV spectrum of a solar flare.
 
 An alternative way to get constraints on photospheric solar neon abundance is to assume that it is comparable to the average neon abundance derived from other objects in the local solar environment, for example, B-type stars, planetary nebulae, H\ii\ regions, and protoplanetary disks.
In our study we focus on normal B-type stars in the solar neighborhood. They are good indicators for neon abundance since they preserve their pristine abundances and allow to measure the neon composition directly from the observed spectra due to their high effective temperatures.

   The Ne\ione\ lines can be observed in B-type stars with effective temperatures higher than 10\,000~K, since the energy of the lowest excited state of neutral Ne atom is quite high (16.6~eV). If the \Teff\ is higher than 21\,000~K, the lines of both ionization stages, Ne\ione\ and Ne\ii\, appear in stellar spectrum. However, neon lines, which are available for neon abundance determination, are known to subject to NLTE effects. There are several studies in the past, where the NLTE neon abundances were derived from B-type stars in the solar neighborhood: 
   \citet{ 1999ApJ...519..303S,2000MNRAS.318.1264D,2003A&A...408.1065H,2006ApJ...647L.143C,2008ApJ...688L.103P,2008A&A...487..307M, 2010PASJ...62.1239T}.
   The neon abundance ranges from the lowest 7.97$\pm$0.07 \citep{2008A&A...487..307M} to the highest value 8.16$\pm$0.14 \citep{2003A&A...408.1065H}.
   It should be noticed that such relatively low neon abundances in these studies exclude the possibility of a considerably high Ne/O ratio once proposed as a solution to the confronted solar model problem. 
   
   The abundance of neon is an important ingredient not only for 'Solar Model Problem', but it also can help to reconstruct the nucleosynthesis history of our Galaxy by sampling neon abundances at a range of the positions in the Galactic disk. Neon is one of the most important contributors to the opacity at the base of the convective zone, after oxygen and iron. 
   The nucleosynthesis of neon proceeds through the $\alpha$ sequence of nuclear reactions, and as such the abundance of neon in a variety of astrophysical objects is an important test of the theory of stellar evolution. Here, we construct a model atom for Ne\ione\ $-$ Ne\ii\ using the most up-to-date atomic data available so far and analyse all observed lines of Ne\ione\ and Ne\ii\ in high resolution spectra of reference B-type stars. Motivated by the 'Solar Model Problem', we collected the sample stars from the solar neighborhood with well known stellar parameters in order to revise neon abundances using actual atomic data.
   
  The paper is organized as follows. Section \ref{Sect:atom} describes an updated model atom of Ne\ione\ $-$ Ne\ii\ and discusses the departures from LTE in the model atmospheres of B-type stars. In Section \ref{sec:stellar}, we analyse the neon lines observed in B-type stars, determine the neon abundance of the selected stars, make some discussions and compare our results with other studies from the literature. 
  Our conclusions are summarized in Section \ref{Sect:Conclusions}.

\section{NLTE line formation for Ne\ione~--   Ne\ii}\label{Sect:atom}

 \subsection{Model atom and atomic data}\label{subSect:atom}

  Neon is a noble gas with the ground configuration 2$s^2$2$p^6$ and its higher levels are formed by the combination of the 2$s^2$2$p^5$ $^2$P$^o_{1/2,3/2}$ core with an excited valence electron.
  The energy levels of Ne\ione\ are poorly described by the LS coupling and are generally described as $j$[K]$_J^{\pi}$, where $\pi$ is the parity of level, $j$ is the total angular momentum of the core, 
  K is the coupling of $j$  with $l$, and $J$ is the total angular momentum \citep{Sobelman1992}.
  Neon has three stable isotopes, $^{20}$Ne, $^{21}$Ne, and $^{22}$Ne, whose abundances in the naturally occurring element are 90.48$\%$,
  0.27$\%$, and 9.25$\%$, respectively \citep{1998JPCRD..27.1275R}.
  
  {\bf Energy levels.}
  We included the ground state and energy levels of Ne\ione\ up to $n$=5 and $l$=$s, p, d, f, g$, belonging to configurations of the type 2$p^5$-$nl$; 
  Ne\ii\ levels belonging to configurations of the type 2s$^2$2p$^5$, 2s2p$^6$, and 2s$^2$2p$^4$-$nl$ ($n$=3 to 5, $l$=$s, p, d, f$); 
   the lowest seven levels of Ne\iii\ without fine structure and the ground state of Ne\iv.
  For Ne\ione, the different states of the atomic core with the total angular momentum, $j$=1/2, 3/2, were taken into account for all considered levels. 
  The splitting as a result of spin-orbit interaction was included only for the levels with $n \le 3$ plus levels with 2$p^5$-4s configuration (the lowest 31 fine-structure levels). As for Ne\ii, the fine structure splitting was included for all levels with $n \le 4$.
  
\begin{figure*}
 \begin{center}
 \includegraphics[scale=0.5]{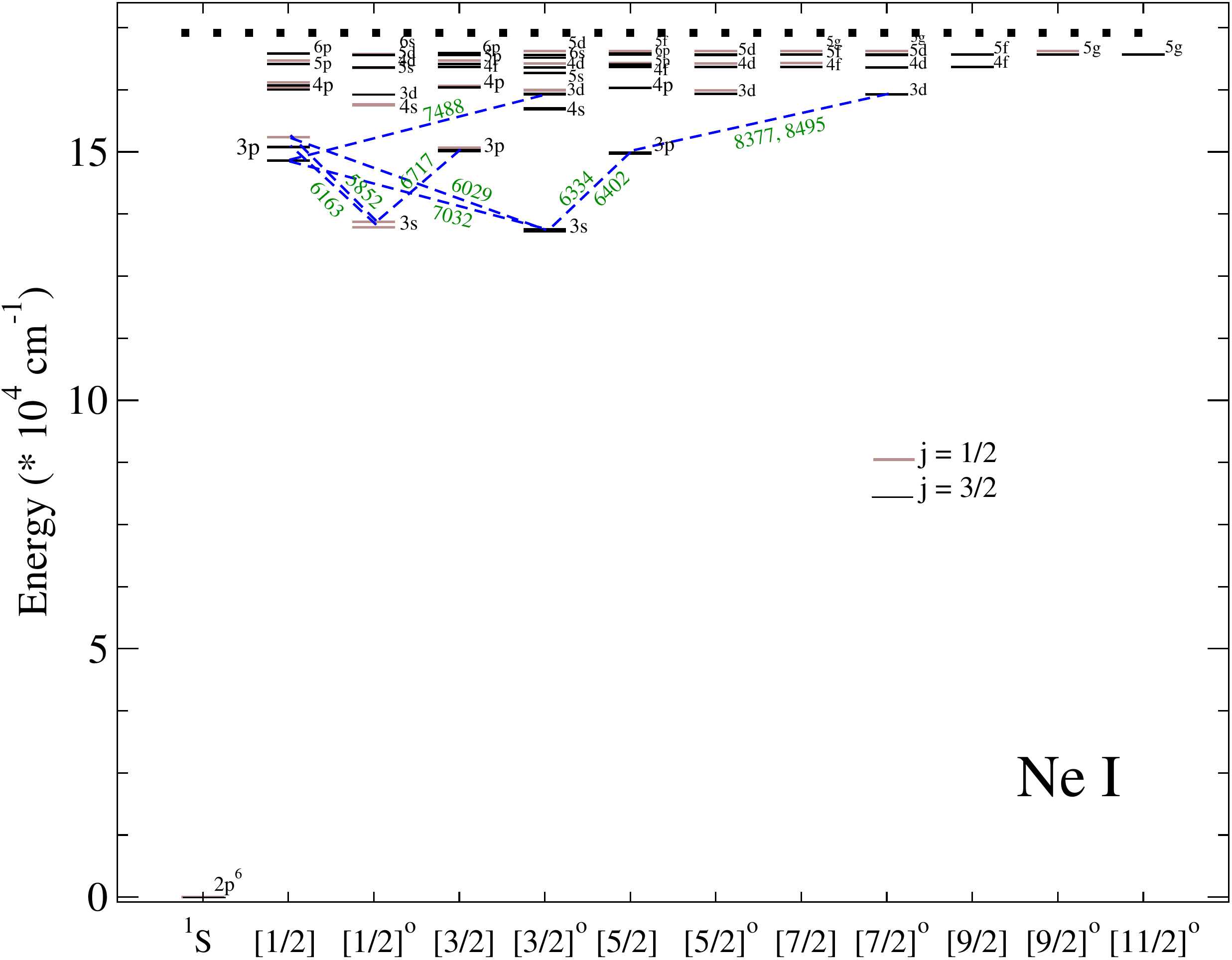}
 \caption{Term diagram for Ne\ione.  The dashed lines indicate the transitions, where the investigated spectral lines arise. The dotted line is the Ne\ione\ ionization threshold. The levels with the  total angular momentum of the atomic core, $j$=1/2, 3/2, are shown separately.} 
 \label{Grot_C2}
 \end{center}
 \end{figure*}

\begin{figure*}
 \begin{center}
 \includegraphics[scale=0.5]{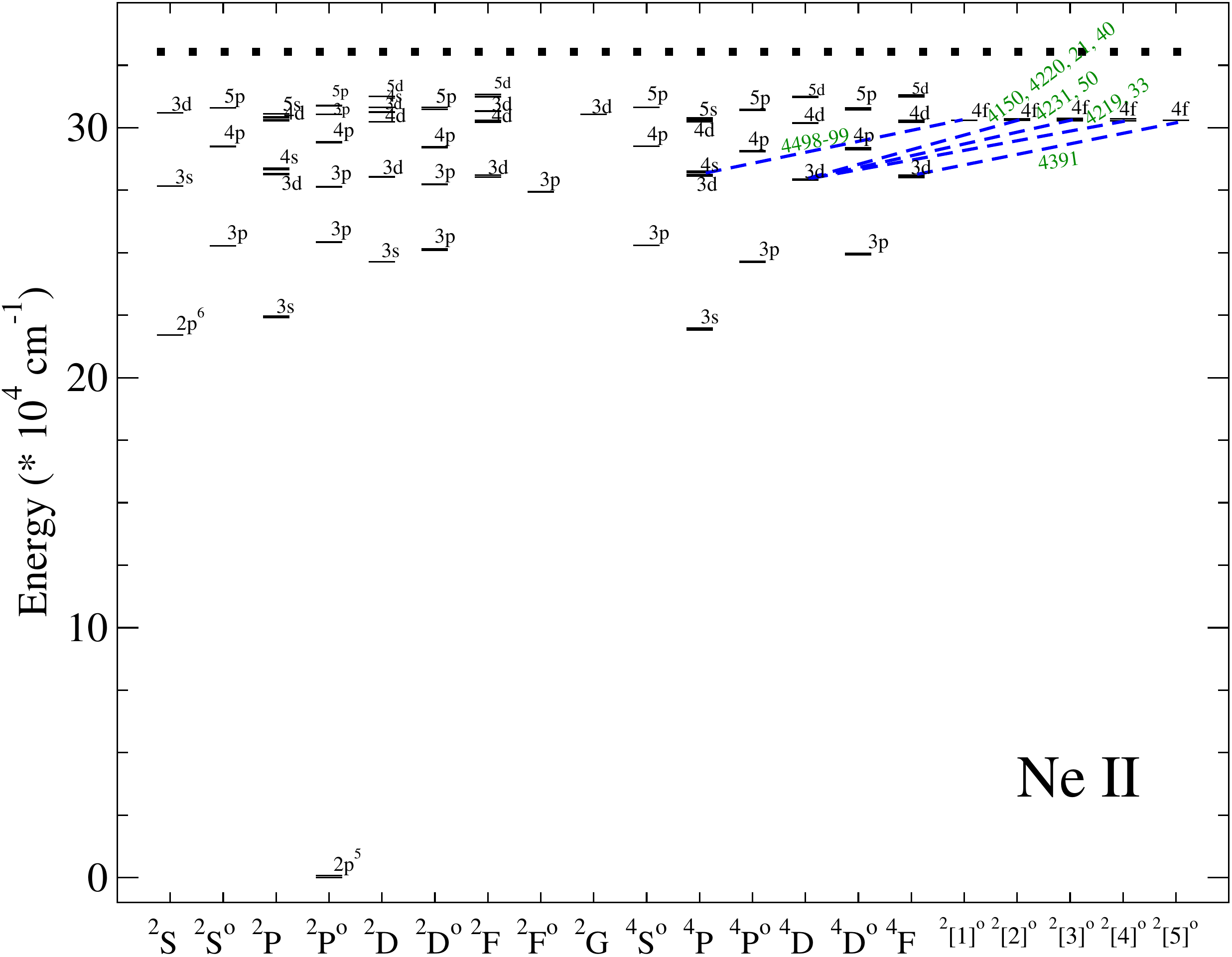}
 \caption{Term diagram for Ne\ii. The dashed lines indicate the transitions, where the investigated spectral lines arise.
  The dotted line is the Ne\ii\ ionization threshold.}
 \label{Grot_C3}
 \end{center}
 \end{figure*}

  Energy levels were taken from the NIST database \footnote{\url{https://physics.nist.gov/PhysRefData/ASD/levels_form.html}}  version 5.7 \citep{NIST_ASD}.
  The term diagrams for Ne\ione\ and Ne\ii\ are shown in Fig.\,\ref{Grot_C2} and Fig.\,\ref{Grot_C3}, respectively. 
 
 \noindent {\bf Radiative and collisional data.} Our model atom includes 446 allowed bound-bound ($b-b$) transitions between Ne\ione\ levels, 1348 $b-b$ transitions between Ne\ii\ levels, and 5 transitions between Ne\iii\ levels.
  The transition probabilities for Ne\ione\ were adopted from \citet{2009PhST..134a4020Z} for transitions between the lowest 31 fine structure levels and from \citet{1998JPhB...31.5315S}
 for the remaining ones.
 The oscillator strengths for Ne\ii\ and Ne\iii\ were taken from the NIST database ver. 5.7 (where available) and from 
	the Kurucz's  website\footnote{\url{http://kurucz.harvard.edu/atoms/1001/gfemq1001.pos}}. These data were extracted via the facilities of the VALD3\footnote{\url{http://vald.astro.uu.se/~vald3/php/vald.php}} database \citep{2015PhyS...90e4005R}. 
 
 We used the effective collision strengths from \citet{2012PhRvA..86b2717Z, 2012PhRvA..85f2710Z} for the transitions connecting the 31 lowest fine structure levels of Ne\ione\ and 
R-matrix electron impact excitation calculations from \citet{2007JPhB...40.2969W} for the transitions connecting the 80 lowest fine structure levels of Ne\ii.
 For the remaining transitions, for which the R-matrix electron impact excitation calculations are not available and for transitions between Ne\iii\ levels, the formula of \citet{1962ApJ...136..906V} was applied for allowed transitions and the effective collision strength $\Omega_{ij}$ = 1 for the forbidden ones.
 The data for Ne\ione\ were provided by Oleg Zatsarinny in an electronic format, while the data for Ne\ii\ were taken from the archives of APAP (Atomic Processes for Astrophysical Plasmas) Network \footnote{\url{http://www.apap network.org/}}. 
  
 The photo ionization cross sections for the levels of Ne\ione,  
Ne\ii, and Ne\iii\ were adopted from the Opacity Project (OP) data base TOPbase (TOPbase) \footnote{\url{http://cdsweb.u strasbg.fr/topbase/topbase.html}} \citep{1993BICDS..42...39C, 1989JPhB...22..389L, 1993AAS...99..179H}.
The levels of Ne\ione\ with $n$=3 and 4 originally treated in LS coupling were re-coupled to the $jl$-coupling terms with weights from \citet{1998MNRAS.300L...1S}.

  Ionization by electronic collisions was everywhere treated by using the \citet{Seaton1962} classical path approximation with threshold photoionization cross sections from TOPbase.

 \subsection{Method of calculations}
  
  We used the code \textsc{detail} \citep{detail} based on the method of accelerated $\Lambda$ iteration \citep{rh91}.
  The \textsc{detail} opacity package was updated by \citet{2011JPhCS.328a2015P} by including bound free opacities of neutral and ionized species.
  The calculated departure coefficients, $b_{\rm{i}}$ = $n_{\rm{NLTE}}$ / $n_{\rm{LTE}}$, were used by the code \textsc{synthV\_NLTE} \citep{2016MNRAS.456.1221R} 
  to calculate the synthetic NLTE line profiles. Here, $n_{\rm{NLTE}}$ and $n_{\rm{LTE}}$ are the statistical equilibrium and thermal (Saha Boltzmann) number densities, respectively. 
  The \textsc{binmag} code\footnote{\url{http://www.astro.uu.se/~oleg/binmag.html}} \citep{2018ascl.soft05015K} was used for automatic spectral line fitting and comparison with the observed spectrum. 
  The typical uncertainties in the fitting procedure with observed profile are less than 0.02 dex for weak lines and 0.03 dex for strong lines.
  For consistency with our NLTE studies of C\ione\ -- C\ii\ \citep{2016MNRAS.462.1123A}, Ti\ione\ -- Ti\ii\ \citep{2016MNRAS.461.1000S}, and Ca\ione\ -- Ca\ii\ \citep{2018MNRAS.477.3343S}, 
  here we use exactly the same model atmospheres for 21~Peg, HD~22136, $\pi$ Cet, and $\iota$~Her, calculated with the code \textsc{LLmodels} \citep{2004AA...428..993S}, as in the earlier papers. 
  For the remaining stars, the calculations were performed using plane parallel (1D), chemically homogeneous model atmospheres from the Kurucz's grid\footnote{\url{http://www.oact.inaf.it/castelli/castelli/grids.html}} \citep{2003IAUS..210P.A20C}. 
  Table \ref{tab1} lists the atomic data for Ne\ione\ and Ne\ii\ lines used in the present line formation analysis.  
  For Ne\ione\ lines belonging to the 2p$^5$3p $-$ 2p$^5$3s transitions array, we adopted data of atomic oscillator strengths from
  the experimental measurements deduced using a neon-filled hollow cathode lamp in conjunction with two spectrographs \citep{2015CaJPh..93...80P}. 
		 For 2p$^5$3d $-$ 2p$^5$3p transitions at 8377~\AA\ and 8495~\AA\, we also used experimental transition probabilities measured with the shock tube \citep{1962PhDT.........2D} and renormalized by NIST group \citep{WSG}. These values of transition probabilities are supported by the accurate theoretical calculations of \citet{1998JPhB...31.5315S}.
  For the remaining Ne\ione\ and for Ne\ii\ lines, the oscillator strengths were taken from the NIST Handbook of Chemistry and Physics \citep[][- Ne\ione]{8043TP}
		and from Kurucz's website. Kurucz's theoretical calculations agree very well with the theoretical calculations by Breit-Pauli R-matrix method used in the earlier papers on neon abundance analyses (see Section~\ref{sect:Comp}).
   Stark collisional data were adopted from the Kurucz's website. 
  
  In the Table~\ref{tabgf}, we present a comparison of the oscillator strengths taken from three sources: Piracha15 -- experimental measurements from \citet{2015CaJPh..93...80P}; NIST -- the oscillator strengths from the NIST database  (ver 5.7); and FF04 -- Breit-Pauli calculations from \citet{2004ADNDT..87....1F}.  This Table demonstrates that the new experimental oscillator strengths agree well with the theoretical 
calculations used in the previous works for neon abundance determinations. It may guarantee the correct comparison of our results with the previous determinations in Sect.~\ref{sect:Comp}.

   \begin{deluxetable*}{lccrclccrc}
   \tablecaption{Lines of Ne\ione\ and Ne\ii\ used in the analysis. \label{tab1}}
   \tabletypesize{\scriptsize}
   \tablehead{
   \colhead{$\lambda$} & \colhead{Transition} & \colhead{\Eexc} & \colhead{log~$gf$} & \colhead{Ref.} & \colhead{$\lambda$} & \colhead{Transition} & \colhead{\Eexc} & \colhead{log~$gf$} & \colhead{Ref.}  \\
   \colhead{\AA\,}     & \colhead{}           & \colhead{eV}    & \colhead{}     & \colhead{}  & \colhead{\AA\,}     & \colhead{}           & \colhead{eV}    & \colhead{}     & \colhead{}   
   }
   \colnumbers
   \startdata                                                                                                                     
    {\bf Ne\ione\ } &                                                                         &         &            &       &{\bf  Ne\ii\ }   &                                              &         &            &         \\
    5852.480  &  ($^2$P$_{1/2}^o$)3s [1/2]$_1^o$ $-$ ($^2$P$_{1/2}^o$)3p  [1/2]$_0$      &  16.84  &   $-$0.455 & \citet{2015CaJPh..93...80P} & 4150.690  &  3d $^4$D$_{1/2}$  $-$ 4f $^2$[2]$_{3/2}^o$   &  34.64  & $-$0.060 &  Kurucz   \\                
    6029.996  &  ($^2$P$_{3/2}^o$)3s [3/2]$_1^o$ $-$ ($^2$P$_{1/2}^o$)3p  [1/2]$_1$      &  16.67  &   $-$0.967 & \citet{2015CaJPh..93...80P} & 4219.367  &  3d $^4$D$_{7/2}$  $-$ 4f $^2$[4]$_{7/2}^o$   &  34.60  &  $-$0.490 &  Kurucz   \\                
    6074.337  &  ($^2$P$_{3/2}^o$)3s [3/2]$_1^o$ $-$ ($^2$P$_{3/2}^o$)3p  [1/2]$_0$      &  16.67  &   $-$0.502 & \citet{2015CaJPh..93...80P} & 4219.745  &  3d $^4$D$_{7/2}$  $-$ 4f $^2$[4]$_{9/2}^o$   &  34.60  &  ~0.714  &  Kurucz   \\                
    6096.163  &  ($^2$P$_{3/2}^o$)3s [3/2]$_1^o$ $-$ ($^2$P$_{1/2}^o$)3p  [3/2]$_2$      &  16.67  &   $-$0.327 & \citet{2015CaJPh..93...80P} & 4220.894  &  3d $^4$D$_{5/2}$  $-$ 4f $^2$[2]$_{5/2}^o$   &  34.61  &  $-$0.106 &  Kurucz    \\                
    6143.062  &  ($^2$P$_{3/2}^o$)3s [3/2]$_2^o$ $-$ ($^2$P$_{3/2}^o$)3p  [3/2]$_2$      &  16.61  &   $-$0.105 & \citet{2015CaJPh..93...80P} & 4221.086  &  3d $^4$D$_{5/2}$  $-$ 4f $^2$[2]$_{3/2}^o$   &  34.61  &  $-$0.778 &  Kurucz  \\                
    6163.593  &  ($^2$P$_{1/2}^o$)3s [1/2]$_0^o$ $-$ ($^2$P$_{1/2}^o$)3p  [1/2]$_1$      &  16.71  &   $-$0.625 & \citet{2015CaJPh..93...80P} & 4231.533  &  3d $^4$D$_{5/2}$  $-$ 4f $^2$[3]$_{5/2}^o$   &  34.61  &  $-$0.115 &  Kurucz   \\                
    6217.281  &  ($^2$P$_{3/2}^o$)3s [3/2]$_2^o$ $-$ ($^2$P$_{3/2}^o$)3p  [3/2]$_1$      &  16.61  &   $-$0.921 & \citet{2015CaJPh..93...80P} & 4231.636  &  3d $^4$D$_{5/2}$  $-$ 4f $^2$[3]$_{7/2}^o$   &  34.61  &   ~0.200  &  Kurucz   \\                
    6266.495  &  ($^2$P$_{1/2}^o$)3s [1/2]$_0^o$ $-$ ($^2$P$_{1/2}^o$)3p  [3/2]$_1$      &  16.71  &   $-$0.375 & \citet{2015CaJPh..93...80P} & 4239.919  &  3d $^4$D$_{3/2}$  $-$ 4f $^2$[2]$_{5/2}^o$   &  34.63  &   $-$0.564 &  Kurucz   \\                
    6304.789  &  ($^2$P$_{3/2}^o$)3s [3/2]$_1^o$ $-$ ($^2$P$_{3/2}^o$)3p  [3/2]$_2$      &  16.67  &   $-$0.759 & \citet{2015CaJPh..93...80P} & 4240.105  &  3d $^4$D$_{3/2}$  $-$ 4f $^2$[2]$_{3/2}^o$   &  34.63  &   $-$0.054 &  Kurucz   \\                
    6334.427  &  ($^2$P$_{3/2}^o$)3s [3/2]$_2^o$ $-$ ($^2$P$_{3/2}^o$)3p  [5/2]$_2$      &  16.61  &   $-$0.272 & \citet{2015CaJPh..93...80P} & 4250.646  &  3d $^4$D$_{3/2}$  $-$ 4f $^2$[3]$_{5/2}^o$   &  34.63  &   ~0.132 &  Kurucz   \\                
    6382.991  &  ($^2$P$_{3/2}^o$)3s [3/2]$_1^o$ $-$ ($^2$P$_{3/2}^o$)3p  [3/2]$_1$      &  16.67  &   $-$0.249 & \citet{2015CaJPh..93...80P} & 4391.991  &  3d $^4$F$_{9/2}$  $-$ 4f $^2$[5]$_{11/2}^o$  &  34.73  &  ~0.918 &  Kurucz   \\                
    6402.248  &  ($^2$P$_{3/2}^o$)3s [3/2]$_2^o$ $-$ ($^2$P$_{3/2}^o$)3p  [5/2]$_3$      &  16.61  &   ~0.341   & \citet{2015CaJPh..93...80P} & 4391.994  &  3d $^4$F$_{9/2}$  $-$ 4f $^2$[5]$_{9/2}^o$   &  34.73  &   $-$0.821 &  Kurucz   \\                
    6506.527  &  ($^2$P$_{3/2}^o$)3s [3/2]$_1^o$ $-$ ($^2$P$_{3/2}^o$)3p  [5/2]$_2$      &  16.67  &   $-$0.023 & \citet{2015CaJPh..93...80P} & 4412.591  &  3d $^4$F$_{9/2}$  $-$ 4f $^2$[4]$_{9/2}^o$   &  34.73  &    ~0.050 &  Kurucz    \\
    6598.952  &  ($^2$P$_{1/2}^o$)3s [1/2]$_1^o$ $-$ ($^2$P$_{1/2}^o$)3p  [1/2]$_1$      &  16.84  &   $-$0.365 & \citet{2015CaJPh..93...80P} &           &                                               &         &            &         \\
    6717.043  &  ($^2$P$_{1/2}^o$)3s [1/2]$_1^o$ $-$ ($^2$P$_{1/2}^o$)3p  [3/2]$_1$      &  16.84  &   $-$0.371 & \citet{2015CaJPh..93...80P} &           &                                               &         &            &        \\
    7032.412  &  ($^2$P$_{3/2}^o$)3s [3/2]$_2^o$ $-$ ($^2$P$_{3/2}^o$)3p  [1/2]$_1$      &  16.61  &   $-$0.248 & \citet{2015CaJPh..93...80P} &           &                                               &         &            &   \\                
    7245.166  &  ($^2$P$_{3/2}^o$)3s [3/2]$_1^o$ $-$ ($^2$P$_{3/2}^o$)3p  [1/2]$_1$      &  16.67  &   $-$0.545 & \citet{2015CaJPh..93...80P} &           &                                               &         &            &  \\  
    7535.774  &  ($^2$P$_{3/2}^o$)3p [1/2]$_1$   $-$ ($^2$P$_{3/2}^o$)3d  [1/2]$_1^o$    &  18.38  &   ~0.040   & \citet{8043TP}         &           &                                               &         &            &  \\ 
    8377.607  &  ($^2$P$_{3/2}^o$)3p [5/2]$_3$   $-$ ($^2$P$_{3/2}^o$)3d  [7/2]$_4^o$    &  18.55  &   ~0.680   &\citet{1962PhDT.........2D} &           &                                               &         &            &  \\ 
    8495.359  &  ($^2$P$_{3/2}^o$)3p [5/2]$_2$   $-$ ($^2$P$_{3/2}^o$)3d  [7/2]$_3^o$    &  18.57  &   ~0.432   & \citet{1962PhDT.........2D} &           &                                               &         &            &  \\ \hline 
   \enddata                                                                                                                                                                                 
   \end{deluxetable*}

    \begin{deluxetable*}{lccc}
   \tablecaption{Comparison of the Oscillator strengths, log~$gf$, for several Ne\ione\ lines \label{tabgf}}
   \def\arraystretch{1.0}
   \setlength{\tabcolsep}{2pt} 
   \tablewidth{20pt}
   \tablehead{
   \colhead{$\lambda$ (\AA)} & \colhead{Piracha15}   & \colhead{NIST}  & \colhead{FF04}    
   }
   \startdata                                                                                               
    5852.480  &   $-$0.455   &   $-$0.500   &   --       \\
    6029.996  &   $-$0.967   &   $-$1.037   &   $-$1.026 \\
    6074.337  &   $-$0.502   &   $-$0.477   &   $-$0.473 \\
    6096.163  &   $-$0.327   &   $-$0.297   &   $-$0.272 \\
    6143.062  &   $-$0.105   &   $-$0.098   &   $-$0.070 \\
    6163.593  &   $-$0.625   &   $-$0.603   &   $-$0.598 \\
    6217.281  &   $-$0.921   &   $-$0.955   &   $-$0.943 \\
    6266.495  &   $-$0.375   &   $-$0.357   &   $-$0.331 \\
    6304.789  &   $-$0.759   &   $-$0.906   &   $-$0.873 \\
    6334.427  &   $-$0.272   &   $-$0.315   &   $-$0.277 \\                                         
    6382.991  &   $-$0.249   &   $-$0.230   &   $-$0.205 \\                                         
    6402.248  &   ~0.341     &   ~0.3456    &     ~0.365 \\                                     
    6506.527  &   $-$0.023   &   $-$0.021   &   $-$0.002 \\                                         
    6598.952  &   $-$0.365   &   $-$0.342   &   $-$0.316 \\                                         
    6717.043  &   $-$0.371   &   $-$0.356   &   $-$0.346 \\                                                                                  
    7032.412  &   $-$0.248   &   $-$0.228   &   $-$0.222 \\                                         
    7245.166  &   $-$0.545   &   $-$0.622   &   $-$0.620 \\                                         
   \enddata                                                                                                                                                                  
   \tablecomments{{\bf Notes.} Piracha15 -- experimental measurements from \citet{2015CaJPh..93...80P}; NIST -- the oscillator strengths from NIST database (ver. 5.7); FF04 -- Breit-Pauli calculations from \citet{2004ADNDT..87....1F} }
   \end{deluxetable*}

 \subsection{Departures from LTE for Ne\ione\ and Ne\ii\ lines in B-type star}\label{Sect:departure}

 \begin{figure*}
 \begin{minipage}{170mm}
 \begin{center}
 \parbox{0.3\linewidth}{\includegraphics[scale=0.25]{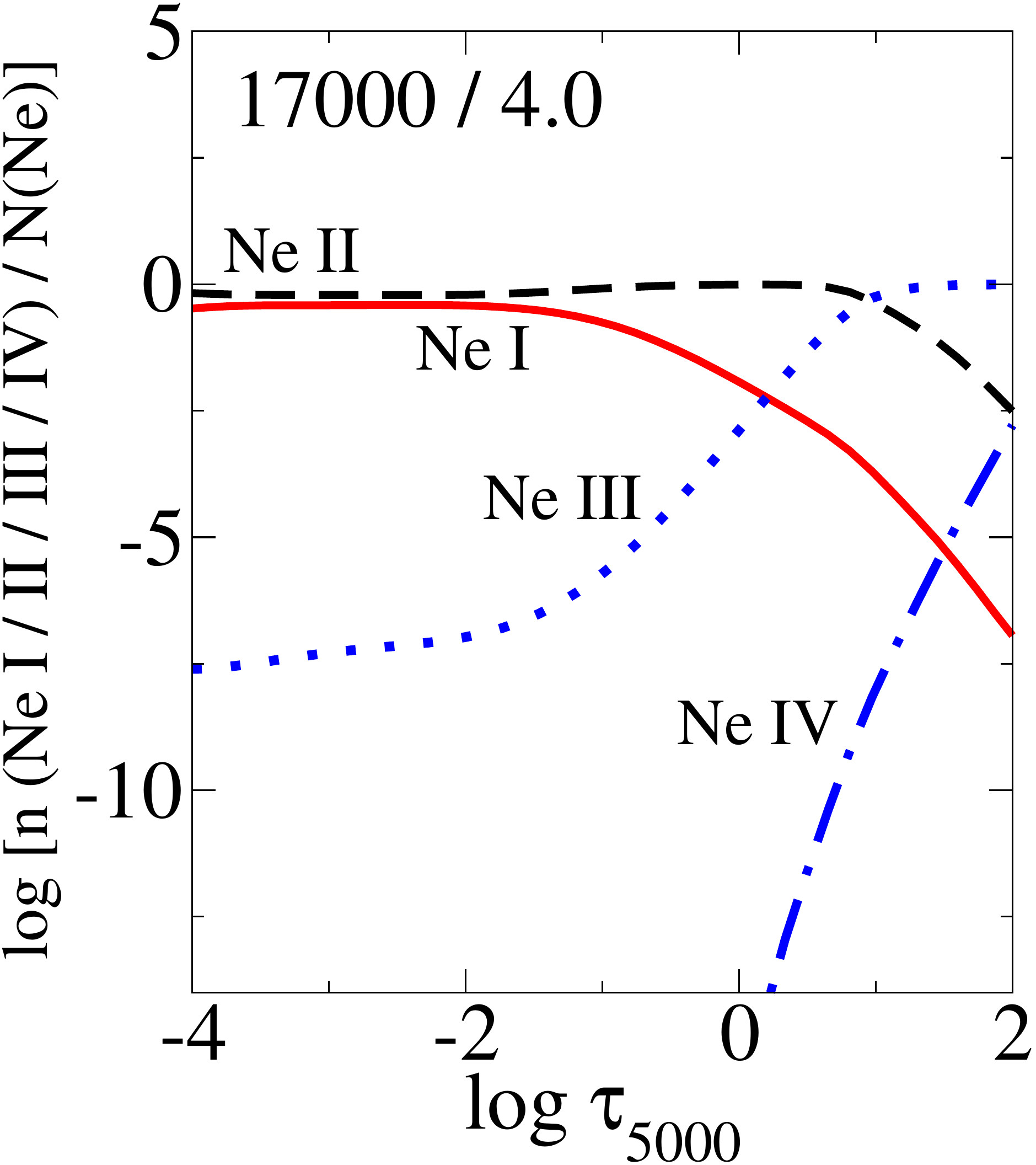}\\
 \centering}
 \parbox{0.3\linewidth}{\includegraphics[scale=0.25]{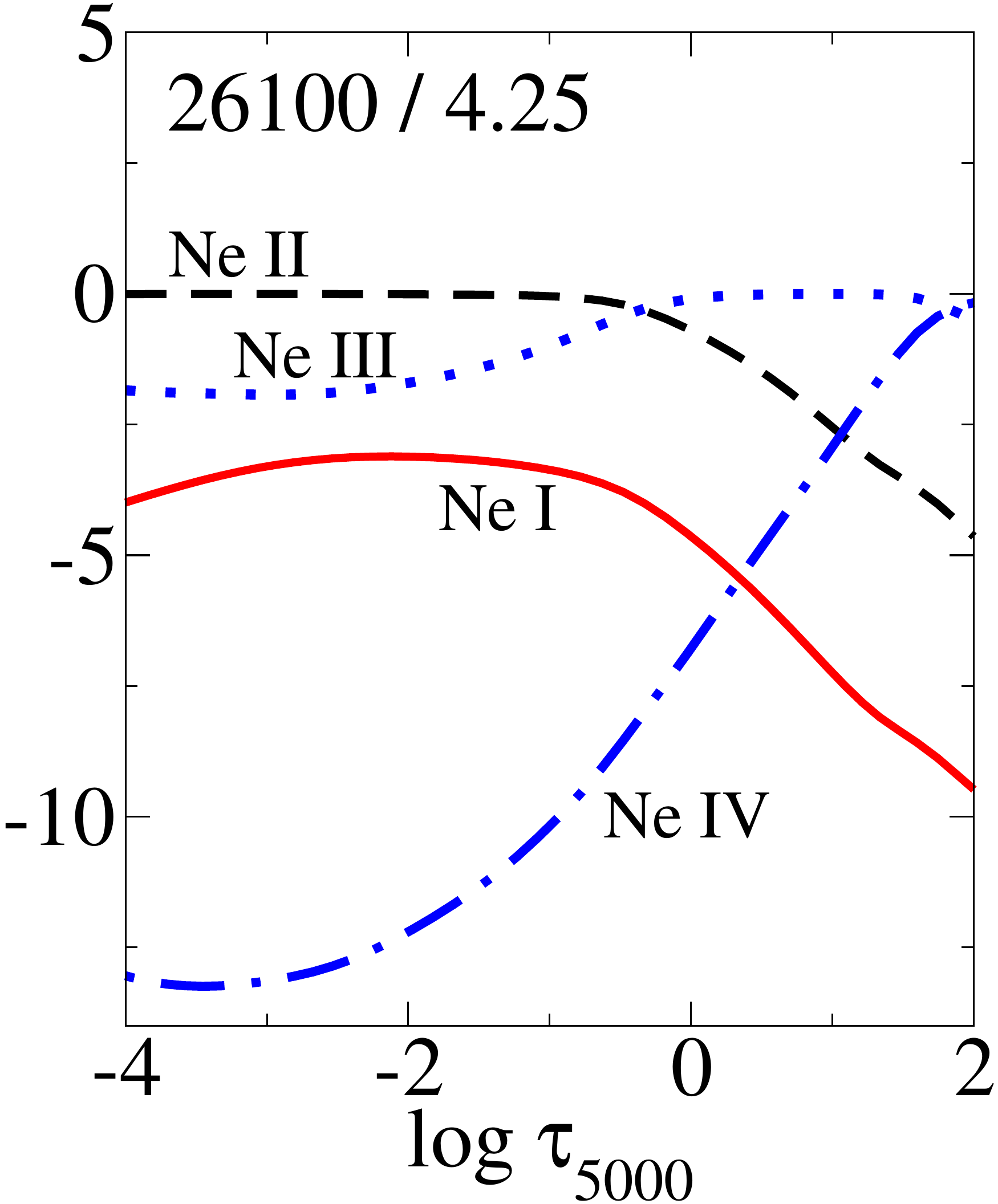}\\
 \centering}
 \hspace{1\linewidth}
 \hfill
 \\[0ex]
 \caption{Ionization fractions of Ne\ione, Ne\ii, Ne\iii, and Ne\iv\ as a function of $\log \tau_{5000}$ in the model atmospheres of 17000 / 4.0 and 26100 / 4.25. Everywhere log~$\epsilon_{\rm Ne}$ = 8.05 and V$_{mic}$ = 5~\kms. }
 \label{balance}
 \end{center}
 \end{minipage}
 \end{figure*} 
 
 \begin{figure*}
  \begin{minipage}{175mm}
  \parbox{0.33\linewidth}{\includegraphics[scale=0.2]{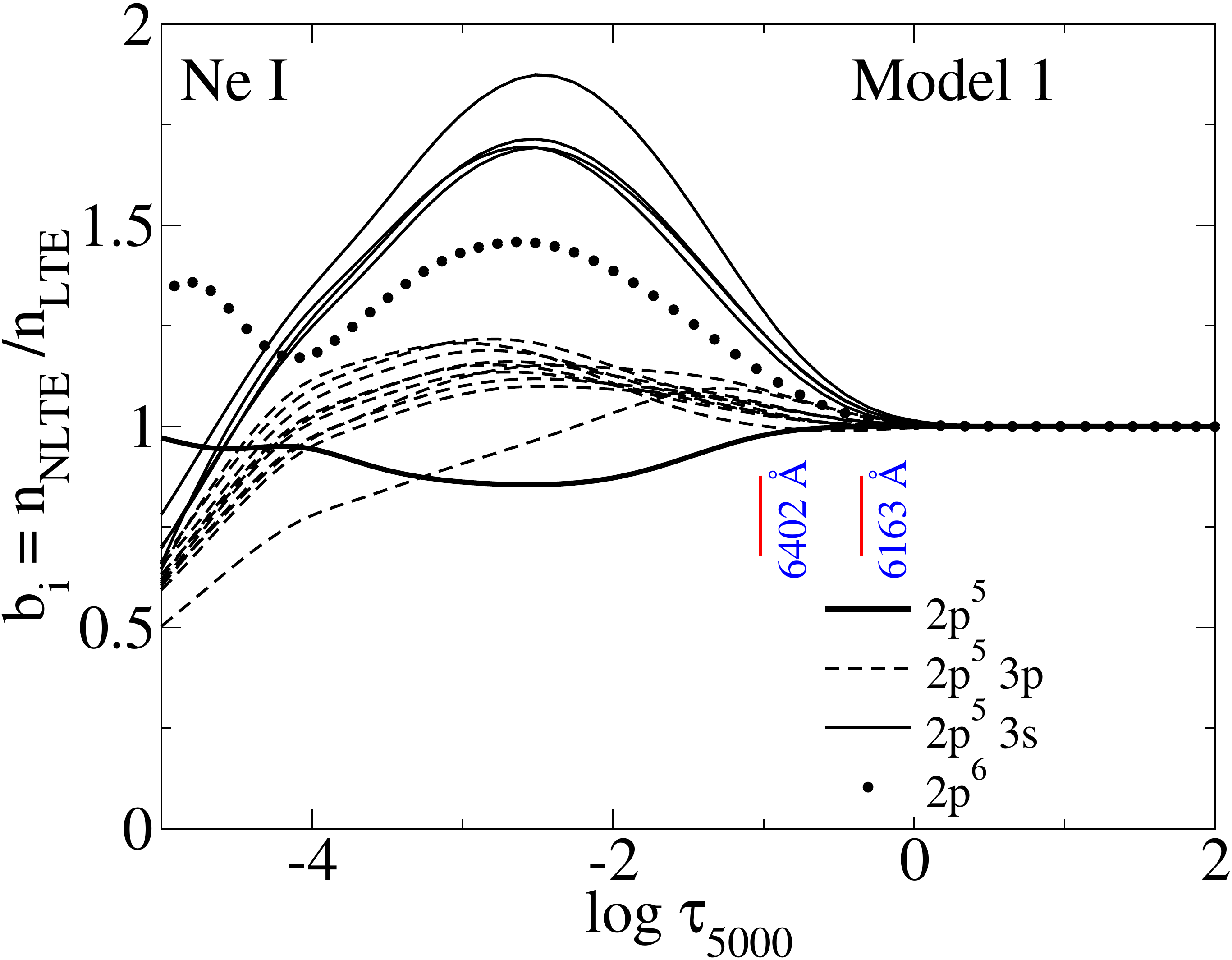}\\
 \centering}
 \parbox{0.33\linewidth}{\includegraphics[scale=0.2]{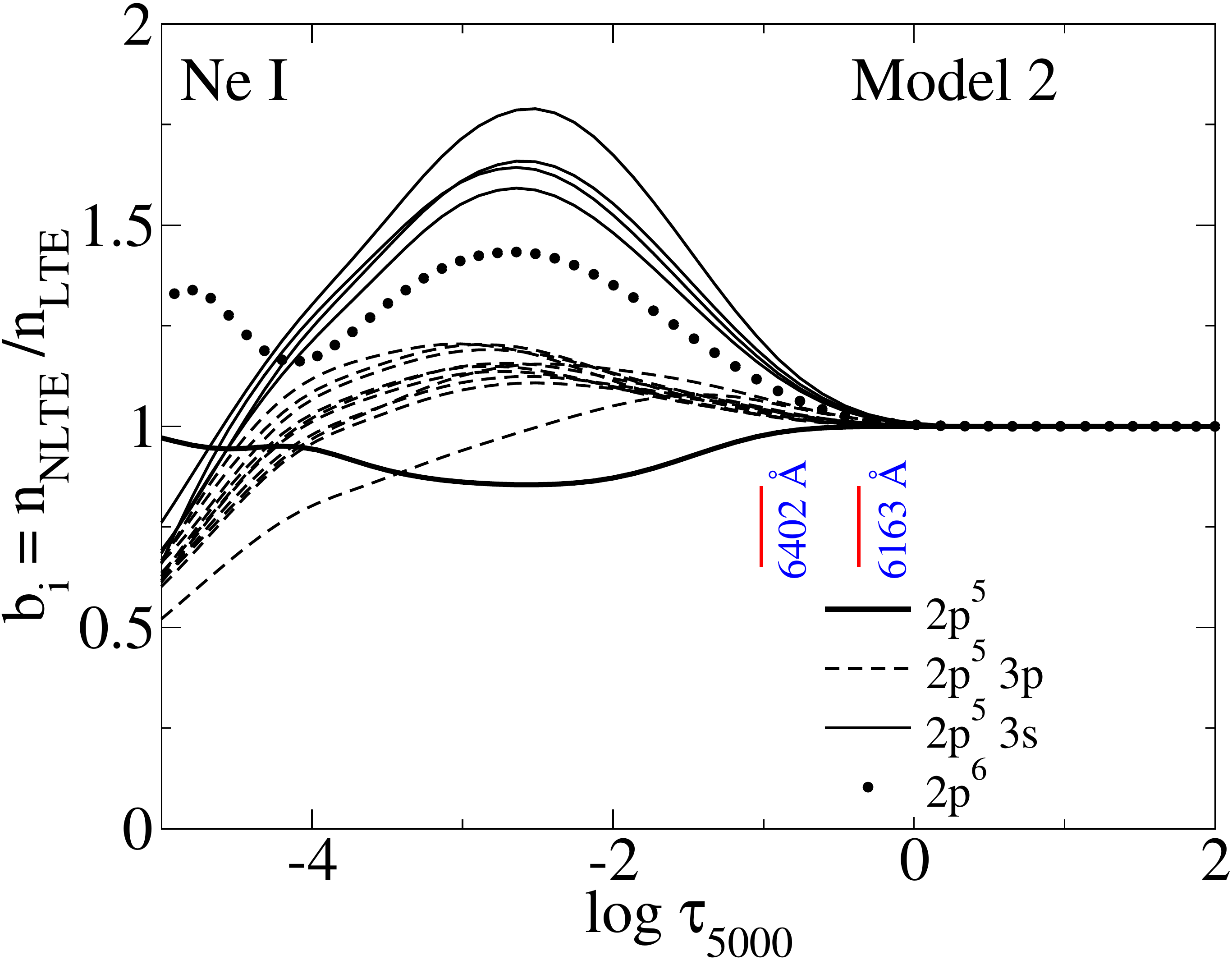}\\
 \centering}
 \parbox{0.33\linewidth}{\includegraphics[scale=0.2]{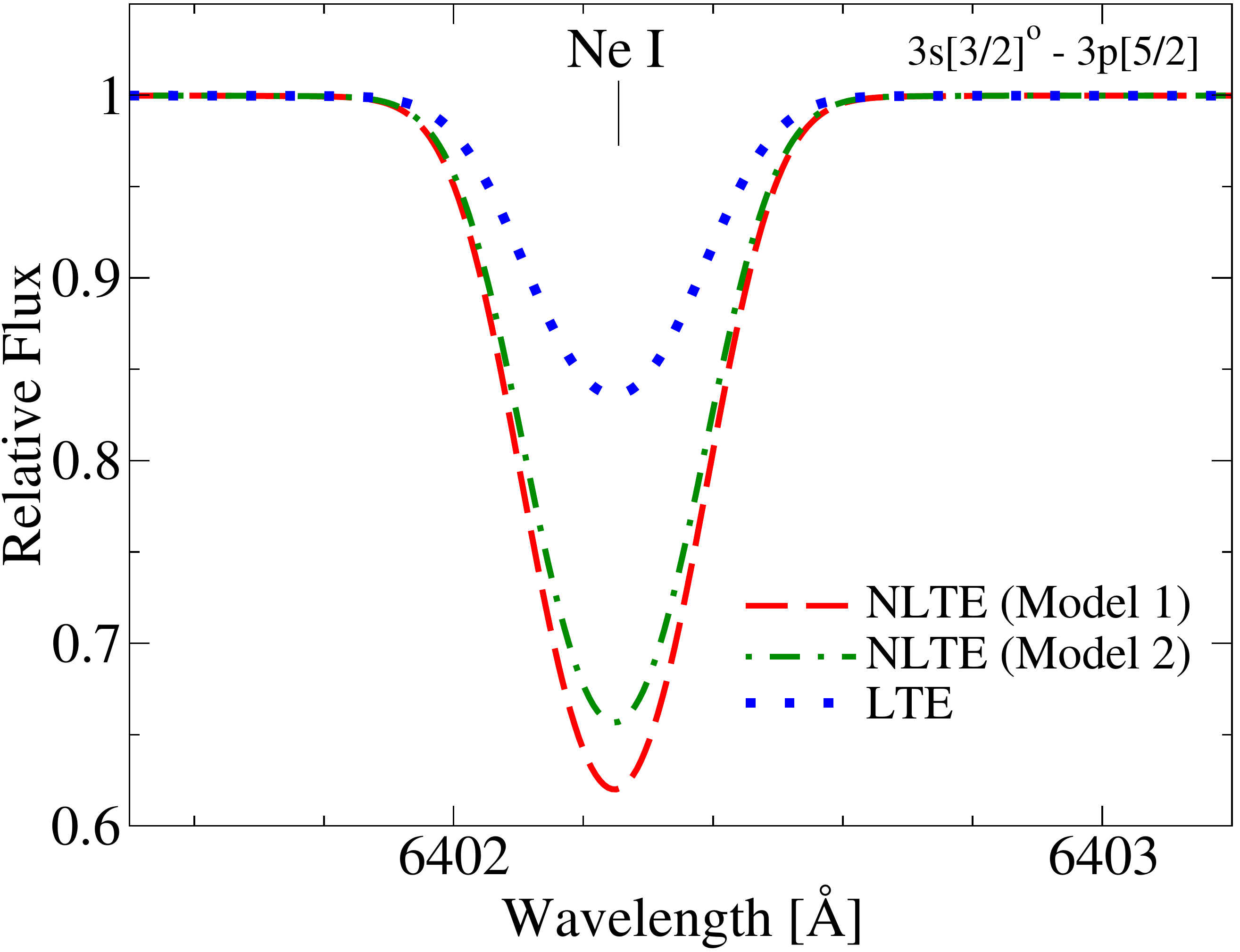}\\
 \centering}
 \hspace{1\linewidth}
 \hfill
 \\[0ex]
 \caption{Departure coefficients as a function of $\log \tau_{5000}$ for the Ne\ione\ levels, calculated with two model atoms in the model atmosphere with \Teff /log~$g$ = 17\,000 / 4.0. The wavelengths and vertical marks indicate the location of optical
depth unity for the lines under consideration. Left panel: Model 1 -- current model atom as described in Sect. \ref{subSect:atom}; Middle panel: Model 2 -- model atom, where electron impact excitation was not included; 
  Right panel: Theoretical line profile of Ne\ione\ at 6402~\AA\ calculated with LTE and NLTE. Everywhere log~$\epsilon_{\rm Ne}$ = 8.08 and V$_{mic}$ = 5~\kms. The theoretical spectra are convolved with an instrumental profile of R~=~50\,000. 
 }
 \label{DC17}
 \end{minipage}
 \end{figure*}

 \begin{figure*}
  \begin{minipage}{175mm}
 \parbox{0.45\linewidth}{\includegraphics[scale=0.27]{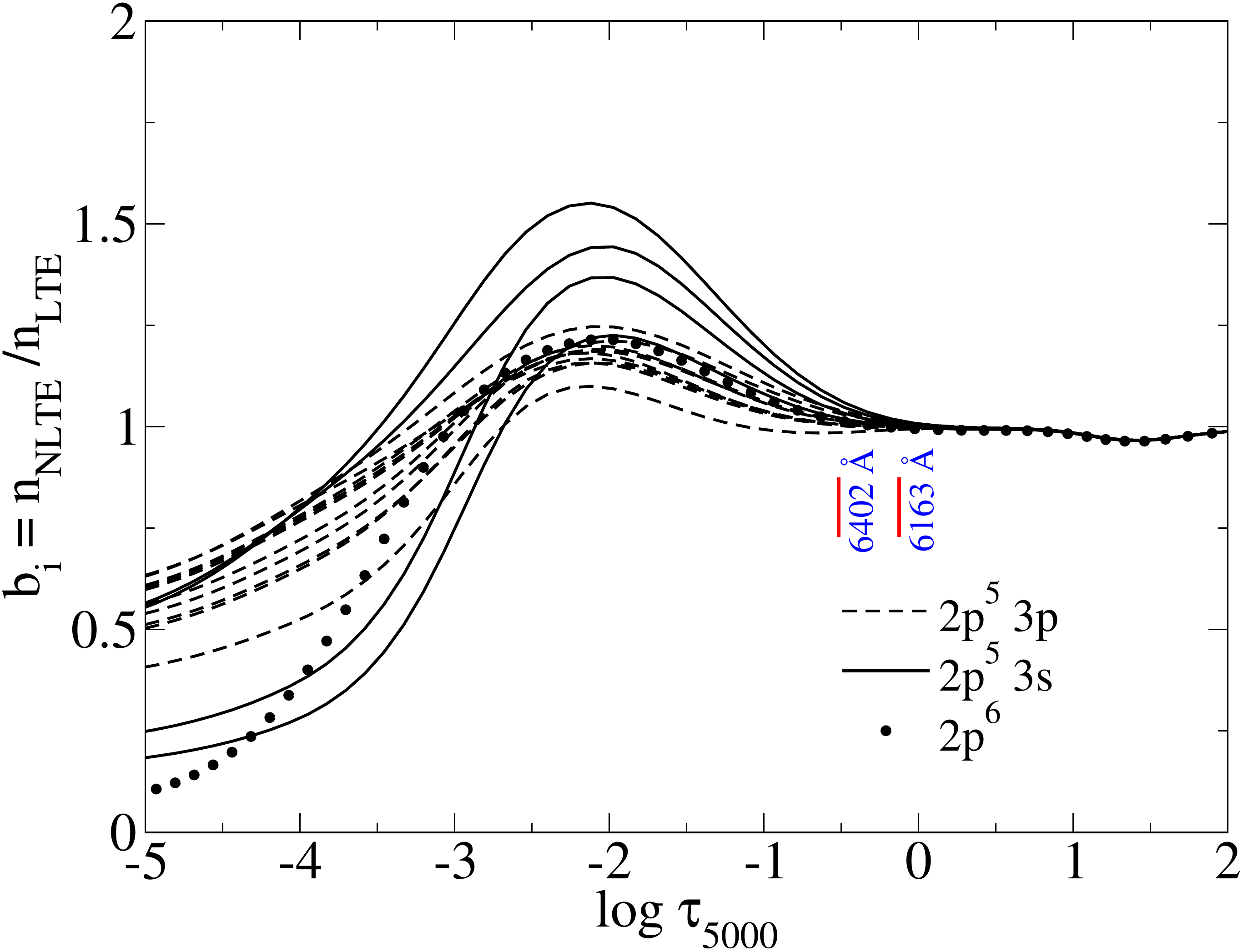}\\
 \centering}
 \parbox{0.45\linewidth}{\includegraphics[scale=0.27]{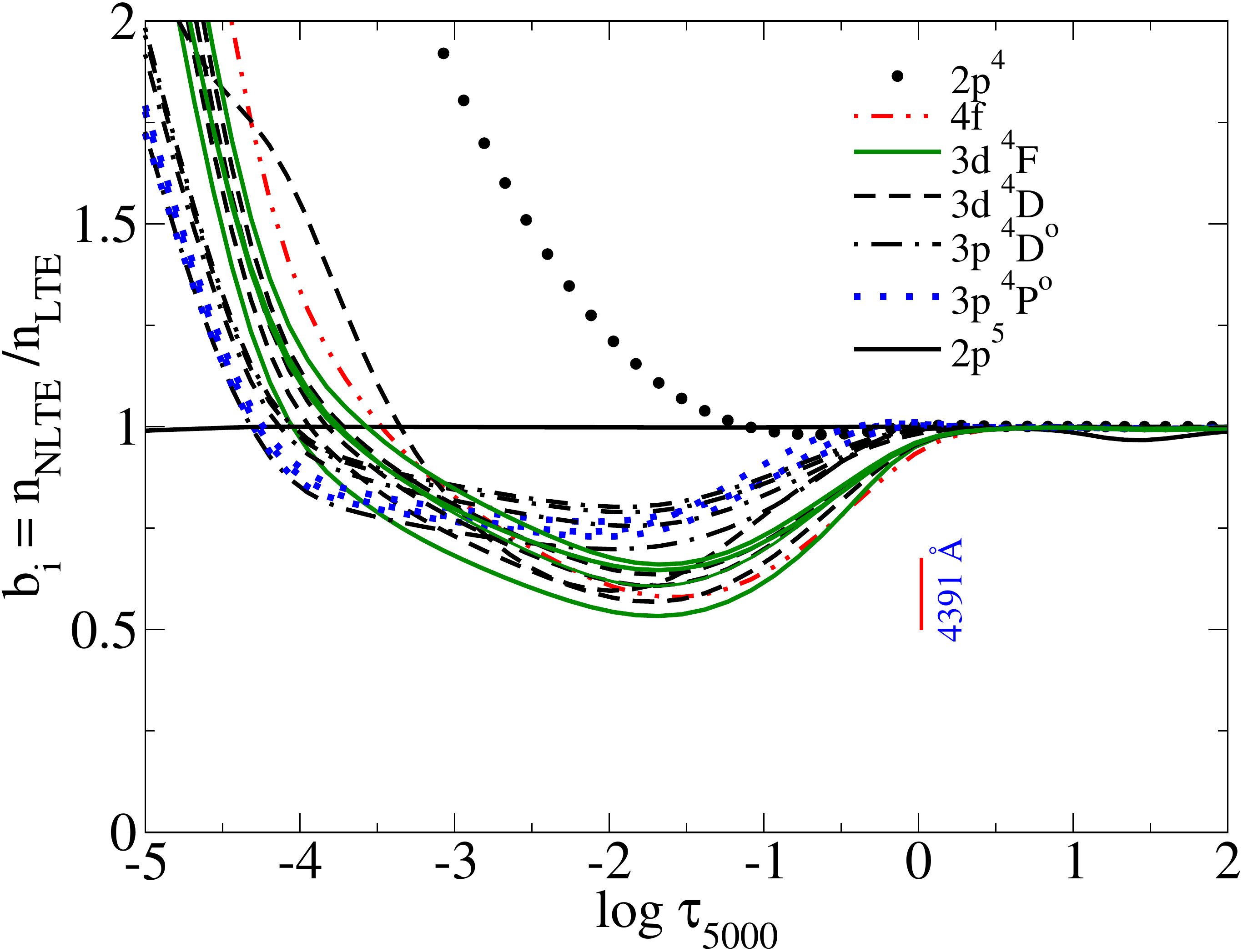}\\
 \centering}
 \hspace{0.00\linewidth}
 \hfill
 \caption{Departure coefficients for the Ne\ione\ (left panel) and Ne\ii\ (right panel) levels as a function of $\log \tau_{5000}$ in the model atmospheres with \Teff /log~$g$ = 26\,100 / 4.25. The wavelengths and vertical marks indicate the location of optical
depth unity for the lines under consideration. Everywhere log~$\epsilon_{\rm Ne}$ = 8.05 and V$_{mic}$ = 5~\kms. }
 \label{DC26}
 \end{minipage}
 \end{figure*}

   Figure\,\ref{balance} displays the fraction of different ions Ne\ione, Ne\ii, Ne\iii\ and Ne\iv\ in the model atmospheres with \Teff /log~$g$ = 17\,000 / 4.0 and 26\,100 / 4.25.  
   In the model 17\,000 / 4.0, neon is predominantly in two ionization stages, Ne\ione\ and Ne\ii, outwards log~$\tau \approx$ -2.5. In deeper layers, where  log~$\tau \geq$ -2.5, Ne\ione\ is ionized and the fraction of Ne\ione\ decreases relative to Ne\ii. 
   In the atmospheres with 26\,100 / 4.25, Ne\ii\ is still the dominant stage in the line formation region, 
   with small admixtures of Ne\ione\ and Ne\iii\ (several thousandths). In deeper layers, where log~$\tau \geq 0.5$, 
   Ne\ii\ is ionized and its fraction decreases relative to Ne\iii. 
 In the atmospheres of the stars with \Teff $\geq$ 25\,000~K, higher ionization stages than  Ne\iii, namely Ne\iv\ and Ne\v, become the dominant species in the deeper atmospheric layers \citep{2007ApJS..169...83L}.

    {\bf Ne\ione:} Figure~\ref{DC17} (left panel) and Figure~\ref{DC26} (left panel) show the departure coefficients for the selected levels of Ne\ione\ in the models 17\,000 / 4.0 and 26\,100 / 4.25, respectively. As expected, the departure coefficients are equal to unity deep in the atmosphere, where the gas density is large and collisional processes dominate, enforcing the LTE.
   The ground state (2p$^6$), 3s, and 3p levels of Ne\ione\ are overpopulated in the line formation region 
   due to the recombinations from Ne\ii\ reservoir to highly excited states followed by the cascade transitions to lower levels.
   The appreciable overpopulation of 3s levels occurs mainly due to  the small photoionization cross-sections of these levels. For example, for 3s[3/2]$^o$ for $\lambda_{th}$ = 2507~\AA\ ionization threshold, the photoionization cross-section $\sigma_0$ = 0.09$\cdot$10$^{-18}$ cm$^{-2}$. 
   
   Most of the Ne\ione\ lines observed in B star spectra (presented in Table~\ref{tab1})  belong to the 3s$-$3p transition array, so the NLTE effects are similar for all of them. 
   The NLTE effects for any spectral line can be understood from analysis of the departure coefficients of the lower (b$_l$) and upper (b$_u$) levels at the line formation depths. A NLTE strengthening of lines occurs, if b$_l>$ 1 and/or the line source function is smaller than the Planck function, that is, b$_l>$ b$_u$.
   In the model 17\,000 / 4.0, all lines from the 3s$-$3p transition array (5852, 6029, 6074, 6096, 6143, 6163, 6217, 6266, 6304, 6334, 6382, 6402, 6506, 6598, 6717, 7032, 7245~\AA) have b$_l>$ b$_u$ and b$_l>$ 1 in their formation region $-$1.3 $\le$ log~$\tau \le$ $-$0.3,  which means that for each line from the array the line source function drops below the Planck function, and the line opacity is amplified.
   Both factors lead to significant strengthening of these lines. 
   For example, for the Ne\ione\ 6402~\AA\ line in the models 17\,000 / 4.0 and 26\,100 / 4.25, the NLTE abundance corrections ($\Delta_{\rm NLTE}$ = log~$\epsilon_{\rm NLTE}$ $-$ log~$\epsilon_{\rm LTE}$) are $-$0.55~dex and $-$0.70~dex, respectively.
   What is the critical factor in so large NLTE corrections? 
   \citet{1973ApJ...184..151A} attributed the strong NLTE strengthening of the 3s$-$3p transitions to amplification of the small NLTE departures by the stimulated emission
   correction to the line source function. However, \citet{1999ApJ...519..303S} showed that the significant NLTE strengthening of
   the line (6402~\AA) would result even if the stimulated emission correction to the line source function was ignored. 
    He suggested that the lines of the 3s$-$3p transition array are dominated by the classical NLTE effect of photon losses in the lines leading to an overpopulation of the lower levels.

   The NLTE effects in Ne\ione\ lines become stronger due to collisions with electrons. 
   Figure\,\ref{DC17} presents the departure coefficients for the selected levels of Ne\ione\ in the model atmospheres with 17\,000 / 4.0 with two model atoms: Model 1 as described in Sect.~\ref{subSect:atom} with collisional data from \citet{2012PhRvA..86b2717Z, 2012PhRvA..85f2710Z} (left panel) and Model 2, where the electron impact excitation was not considered (middle panel).
   The inclusion of the electron collisions leads to increasing the NLTE effects, since 3s levels tend to be more populated relative to the case, where electron collisions are excluded (Model 2). 
   For example, the Ne\ione\ 6402~\AA\ line (3s [3/2]$^o$ $-$ 3p [5/2]) is strong and its core forms around $\log \tau \approx  -1.0$, where b$_l>$ 1 and b$_l>$ b$_u$. 
   Both factors lead to the line source function to be smaller than the Planck function, but in Model 1, $S_{\nu} \approx 0.7$B$_{\nu}$, and in Model 2, $S_{\nu} \approx 0.8$B$_{\nu}$. 
   The effect is estimated as log~$\epsilon_{\rm NLTE(Model 1)}$~$-$~log~$\epsilon_{\rm NLTE(Model 2)}$ = $-$0.12 dex (Figure~\ref{DC17}, right panel).

   {\bf Ne\ii:} The departure coefficients for the selected levels of Ne\ii\ in the model 26\,100 / 4.25 are shown on Figure~\ref{DC26} (right panel). 
   Since Ne\ii\ dominates over all atmospheric depths, the Ne\ii\ ground state keeps its LTE populations.
 However, we noticed small departures from the equilibrium value at $1 < \log \tau < 2$. 
This behavior is highly unusual since the inner boundary condition should give b$_i$ = 1 at log~$\tau > 1$. 
Our test calculations show that it is explained by strong photo ionization of Ne\ii\ from its ground state. It is reinforced by 
Fig.\,\ref{balance} (right panel), where the fraction of Ne\ii\ drops below at $\log \tau > 0$.
 The excited levels of Ne\ii\ are underpopulated relative to their LTE populations, outward log~$\tau = 0$, due to spontaneous transitions. 
   The Ne\ii\ 4391~\AA\ line forms in the layers around $\log \tau =  0.05$, where b$_l<$ 1 and b$_l>$ b$_u$. The first factor leads to the line weakening, but the second one leads to the line strengthening.
   As a result, the NLTE abundance correction is quite small $-$0.01~dex.

  \section{Neon lines in the selected stars} \label{sec:stellar}
  
  \subsection{Stellar sample, Observations, and stellar parameters}  

  Our sample includes 24 bright B-type stars with well known atmospheric parameters (Table~\ref{tab_param}).
  We collected B-type stars in a wide range of spectral types from B0 to B9.5, where neon lines are available for measurements. 
  All of these stars show relatively sharp spectral lines (slowly rotating). 
  Atmospheric parameters of 21 B-type stars were adopted from \citet{2009AA...503..945F}, \citet{2011AA...532A...2N}, and \citet{2012AA...539A.143N}.
  For HD~22136, the parameters were adopted from \citet{2013AA...551A..30B}, and, for 134~Tau and $\tau$~Her, from \citet{1993AA...274..335S}. 
                   
\begin{figure*}
  \begin{minipage}{175mm}
 \parbox{0.45\linewidth}{\includegraphics[scale=0.27]{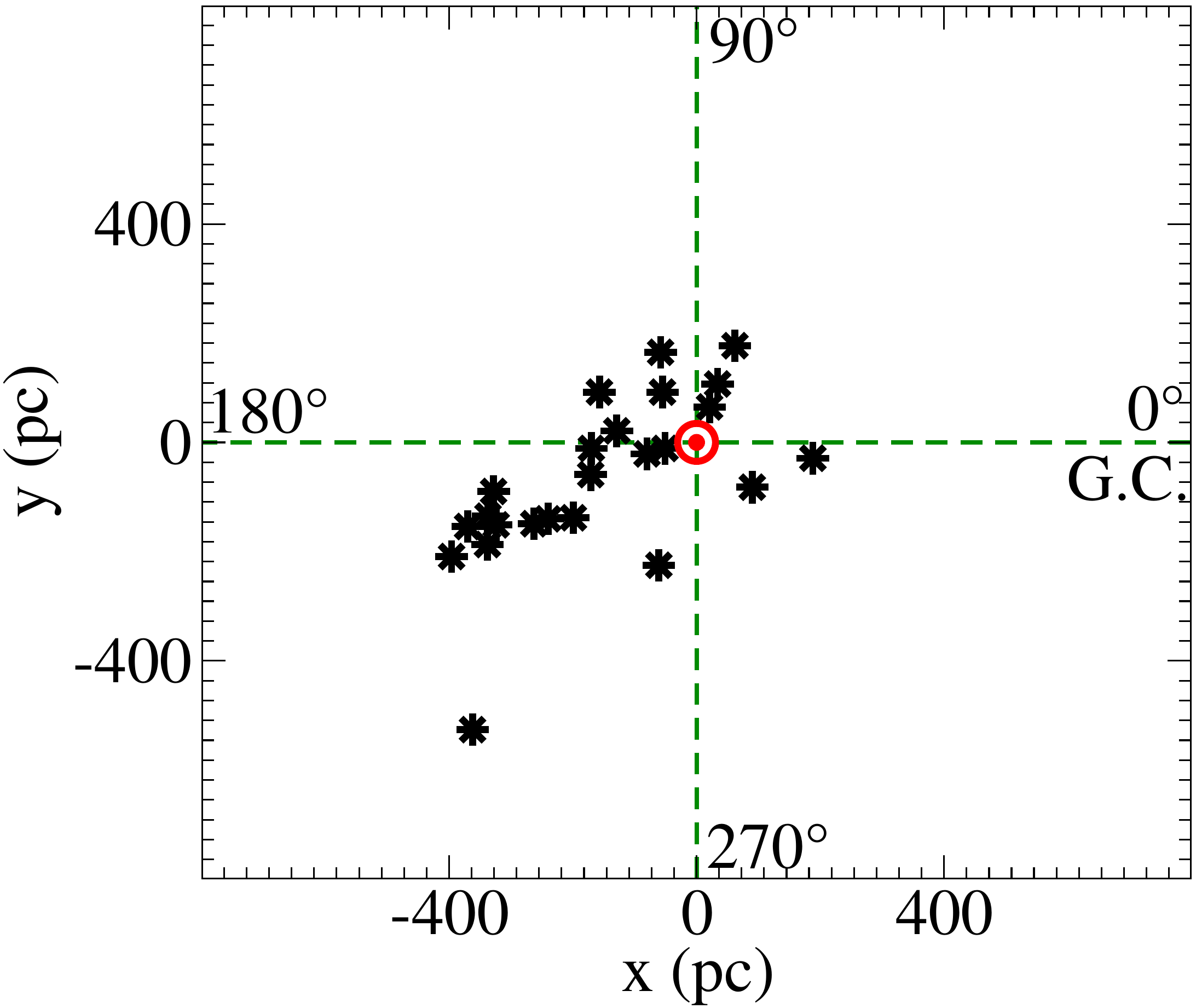}\\
 \centering}
 \parbox{0.45\linewidth}{\includegraphics[scale=0.27]{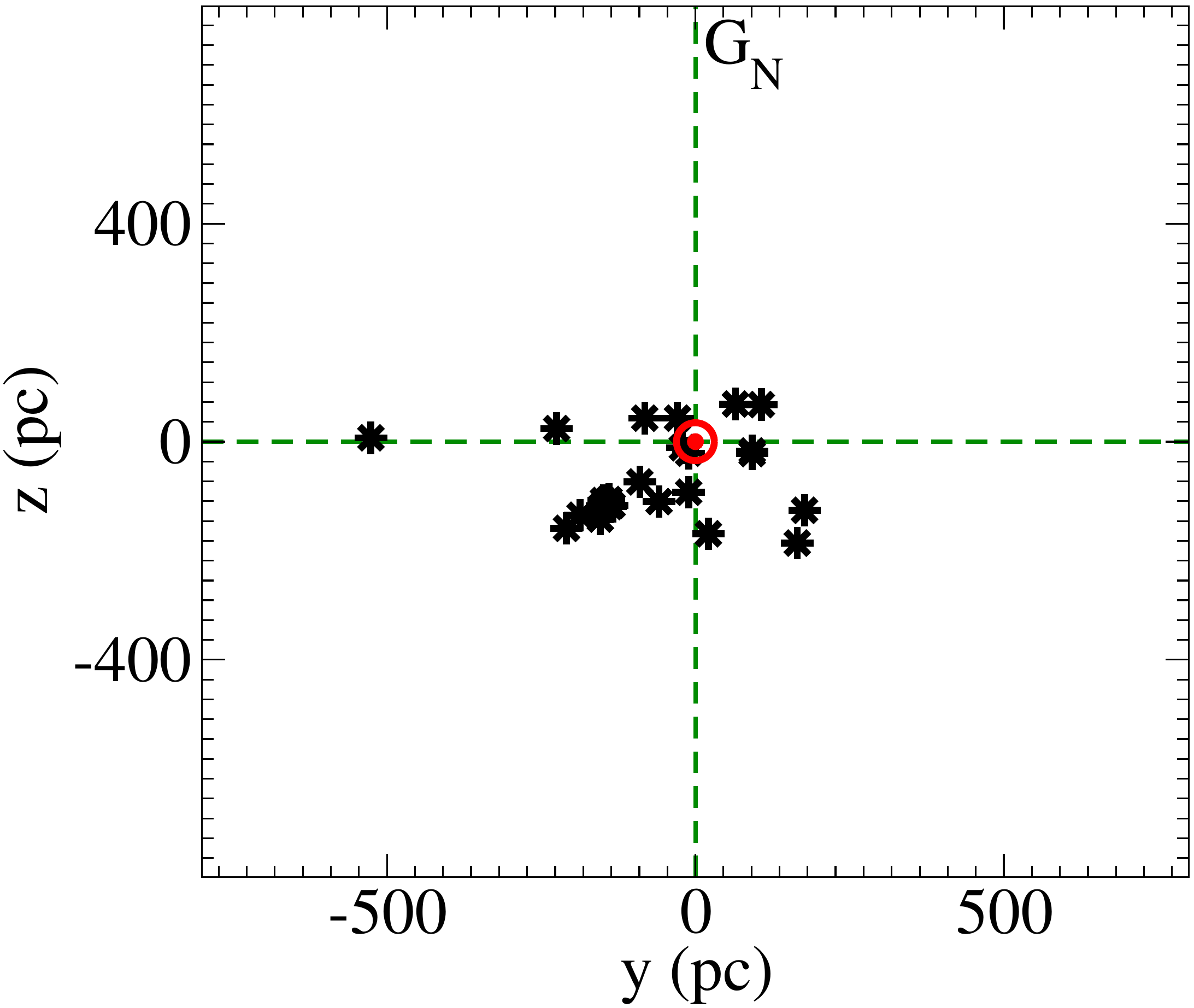}\\
 \centering}
 \hspace{0.00\linewidth}
 \hfill
 \caption{ The positions of the sample stars in the solar neighborhood. Left panel: the projection of stars on the galactic plane. The Galactic centre (G.C.) is in the direction of longitude $l$ = 0$^\circ$.
 Right panel: the projection on the rotational plane.
The Northern Galactic Pole (G$_N$) lies in direction of the top of the panel. The position of the Sun is marked by $\odot$.}
 \label{COOR}
 \end{minipage}
 \end{figure*}

 All the sample stars are from solar neighborhood with the distances from the Sun are no more than $\sim$640 pc.
  The distances for them were calculated from the \textsc{Gaia} parallaxes (Gaia DR2, \citet{2018A&A...616A...1G}). 
  Figure~\ref{COOR} presents the positions of the stars in the Galaxy in two projections, (x;y) and (y;z), where x~=~$d$Cos($b$)Cos($l$), y~=~$d$Cos($b$)Sin($l$), z~=~$d$Sin($b$).

  High resolution ($\it R$ = 65\,000) spectral data 
from the visible to the near IR ranges of our target stars were obtained with the Echelle Spectro Polarimetric Device for the Observation of Stars 
  (ESPaDOnS) \citep{2006ASPC..358..362D} attached to the 3.6 m telescope 
  of the Canada France Hawaii Telescope (CFHT) observatory located on the summit of Mauna Kea,
   Hawaii\footnote{http://www.cfht.hawaii.edu/Instruments/Spectroscopy/Espadons/}. 
  Observations with this spectrograph cover the wavelength region from 3690~\AA\ to 10480 \AA, and we use data from 3855 \AA\ to 9980 \AA\ in the present study. 
  Calibrated intensity spectral data were extracted from the ESPaDOnS archive through Canadian Astronomical Data Centre (CADC). 
  After averaging downloaded individual spectral data of each star, we converted the wavelength scale of spectral data of each star into the laboratory scale using measured wavelengths
  of five He~{\sc i} lines (4471.48, 4713.15, 4921.93, 5015.68, and 5875.62 \AA). 
  Errors in the wavelength measurements are around $\pm$ 3 km s${}^{ -1}$ or smaller. Continuum fitting of each spectral order was carried out using polynomial functions. 
  The signal-to-noise ratios (S/N) measured at the continuum near 5550 \AA\ range from 380 to 1800.

    \begin{deluxetable*}{cllccclccccccc}
    \tablecaption{Atmospheric parameters of the selected stars and sources of the data. \label{tab_param}}
    \tablewidth{0pt}
    \tablehead{  
    \colhead{Number} &\colhead{Star} & \colhead{Name} & \colhead{$l^\circ$}& \colhead{$b^\circ$}& \colhead{V}& \colhead{Sp. T.} & \colhead{ \Teff } & \colhead{log~$g$} & \colhead{$\xi_t$ }  
    & \colhead{$v$~sin~$i$} & \colhead{Ref.} & \colhead{$d_{\rm GAIA}$} & \colhead{S/N}  \\
    \colhead{} &\colhead{} & \colhead{} & \colhead{} & \colhead{} & \colhead{} & \colhead{} & \colhead{ K} & \colhead{CGS}  & \colhead{ \kms } & \colhead{ \kms }   & \colhead{} & \colhead{pc} & \colhead{}  }
    \decimalcolnumbers
    \startdata
    1   & HD~209459 & 21~Peg             & ~70.65  & $-$33.92 & 5.8  & B9.5 V C  &  10\,400  &  3.50   & 0.5     & 4  &  3    &  227  &   600    \\       
    2   & HD~38899  & 134~Tau            & 194.49  & $-$7.57  & 4.9  & B9 IV     &  10\,850  &  4.10   & 1.6     & 30 &  5    &  84   &   490        \\
    3   & HD~22136  &                    & 149.75  & $-$7.03  & 6.8  & B8 V C    &  12\,700  &  4.20   & 1.1     & 15 &  4    &  183  &   500    \\          
    4   & HD~17081  &  $\pi$ Cet         & 191.80  & $-$60.56 & 4.2  & B7 IV E   &  12\,800  &  3.80   & 1.0     & 20 &  3    &  107  &   600     \\ 
    5   & HD~147394 & $\tau$~Her         &  72.48  & 45.03    & 3.9  & B5 IV C   &  15\,000  &  3.95   & 0.0     & 32 &  5    &  97   &   550        \\   
    6   & HD~160762 & $\iota$~Her        & ~72.32  & +31.26   & 3.8  & B3 IV     &  17\,500  &  3.80   & 1.0     & 6  &  1    &  132  &   1820   \\    
    7   & HD~35912  & HR~1820            & 201.88  & $-$17.84 & 6.3  & B2 3 V    &  19\,000  &  4.00   & 2.0     & 15 &  2    &  381  &   650    \\    
    8   & HD~36629  & HIP~26000          & 207.95  & $-$19.52 & 7.6  & B2 V      &  20\,300  &  4.15   & 2.0     & 10 &  2    &  476  &   630    \\   
    9   & HD~35708  & $o$~Tau            & 183.75  & $-$7.17  & 4.8  & B2.5 IV   &  20\,700  &  4.15   & 2.0     & 25 &  1    &  172  &   820     \\   
    10  & HD~3360   & $\zeta$ Cas        & 120.77  & $-$8.91  & 3.6  & B2 IV     &  20\,750  &  3.80   & 2.0     & 20 &  1    &  109  &   1570   \\    
    11  & HD~122980 & $\chi$ Cen         & 317.73  & +19.53   & 4.3  & B2 V      &  20\,800  &  4.22   & 3.0     & 18 &  1    &  129  &   520     \\   
    12  & HD~16582  & $\delta$ Cet       & 170.76  & $-$52.21 & 4.0  & B2 IV     &  21\,250  &  3.80   & 2.0     & 15 &  1    &  213  &   1050    \\   
    13  & HD~29248  & $\nu$ Eri          & 199.31  & $-$31.37 & 3.9  & B2 II     &  22\,000  &  3.85   & 6.0     & 26 &  1    &  212  &   550     \\                     
    14  & HD~886    & $\gamma$ Peg       & 109.43  & $-$46.68 & 2.8  & B2 IV     &  22\,000  &  3.95   & 2.0     & 9  &  1    &  255  &   1110   \\       
    15  & HD~74575  & $\alpha$ Pix       & 254.99  & +5.76    & 3.6  & B1.5 III  &  22\,900  &  3.60   & 5.0     & 11 &  1    &  235  &   800    \\      
    16  & HD~35299  & HR~1781            & 202.67  & $-$19.49 & 5.7  & B2 III    &  23\,500  &  4.20   & 0.0     & 8  &  1    &  425  &   380    \\      
    17  & HD~36959  & HR~1886            & 209.56  & $-$19.72 & 5.5  & B1 V  Ic  &  26\,100  &  4.25   & 0.0     & 12 &  2    &  321  &   860    \\                  
    18  & HD~61068  & HR~2928            & 235.53  & +0.60    & 5.6  & B2 II     &  26\,300  &  4.15   & 3.0     & 14 &  1    &  638  &   560    \\      
    19  & HD~36591  & HR~1861            & 205.13  & $-$18.20 & 5.3  & B2 III    &  27\,000  &  4.12   & 3.0     & 12 &  1    &  377  &   720   \\                
    20  & HD~37042  & $\theta^2$ Ori B   & 209.05  & $-$19.35 & 6.3  & B0.5 V    &  29\,300  &  4.30   & 2.0     & 30 &  2    &  409  &   690   \\    
    21  & HD~36822  & $\phi^1$ Ori       & 195.40  & $-$12.28 & 4.4  & B0.5 III  &  30\,000  &  4.05   & 8.0     & 28 &  1    &  348  &   570   \\    
    22  & HD~34816  & $\lambda$ Lep      & 214.82  & $-$26.23 & 4.2  & B0.5 V    &  30\,400  &  4.30   & 4.0     & 30 &  1    &  270  &   550    \\   
    23  & HD~149438 & $\tau$ Sco         & 351.53  & +12.80   & 2.8  & B0.2 V    &  32\,000  &  4.30   & 5.0     & 4  &  1    &  195  &   1730   \\   
    24  & HD~36512  & $\upsilon$ Ori     & 210.43  & $-$20.98 & 4.6  & B0 V      &  33\,400  &  4.30   & 4.0     & 20 &  1    &  298  &   700    \\      
    \enddata                                                
    \tablecomments{{\bf Note.} References: 
      1 \citet{2012AA...539A.143N}; 2 \citet{2011AA...532A...2N}; 3 \citet{2009AA...503..945F}; 4 \citet{2013AA...551A..30B}; 5 \citet{1993AA...274..335S}
     Visual magnitudes (V) and spectral types (Sp. T.) are extracted from the SIMBAD database. (14) Signal to noise rations (S/N) were measured near 5550 \AA.  }
    \end{deluxetable*}


 \subsection{Testing of the model atom for Ne\ione\ and Ne\ii}

 We selected three stars HD~36959, HD~61068, and HD~36591, in which the neon lines of both ionization stages, Ne\ione\ and Ne\ii, are available.
 For the lines listed in Table~\ref{Netest}, we determined the element abundance under various line-formation assumptions: LTE, and NLTE with three model atoms: Model 1 -- current model atom as described in Sect. \ref{subSect:atom} with electron collisional data for Ne\ione\ from \citet{2012PhRvA..86b2717Z, 2012PhRvA..85f2710Z} for 461 transitions and collisional data for Ne\ii\ from \citet{2007JPhB...40.2969W} for 3078 transitions and formula of \citet{1962ApJ...136..906V} and $\Omega$=1
 for the remaining allowed and forbidden transitions; Model 2, where the electron impact excitation was not included for both Ne\ione\ or Ne\ii; 
 and Model 3, where for allowed and forbidden transitions of the Ne\ione\ and Ne\ii\ the only formula of \citet{1962ApJ...136..906V} and $\Omega$=1 were employed. 
  
 For Ne\ione, the inclusion of the R-matrix electron impact excitation calculations (Model 1) leads to systematically lower abundances compared to the cases in which the electron collisions were not included (Model 2) or 
 the formula of \citet{1962ApJ...136..906V} was used only (Model 3). 
 For Ne\ii\ lines the NLTE effects are quite small, since the Ne\ii\ is the dominant stage and the lines form deeper in atmosphere in these stars.
 The use of the R-matrix electron impact excitation calculations for Ne\ii\ (Model 1) leads to lower abundances compared to LTE case.
 In LTE, the abundance difference between the Ne\ione\ and Ne\ii\ lines, Ne\ione\ $-$~Ne\ii, amounts to 0.45, 0.50, 0.38~dex, for HD~36959, HD~61068, and HD~36591, respectively.
 The best agreement between Ne\ione\ and Ne\ii\ lines is achieved for all stars in the case of the NLTE Model 1, where the abundances from the Ne\ione\ and Ne\ii\ lines are consistent within 0.03~dex for three stars.
 
 Our uncertainties are only limited to the line-to-line scatter, and we did not take into account the uncertainties in atomic data. The real uncertainties of the abundances can be larger than the values deduced from the line-to-line scatter. 
For example, \citet{1999ApJ...519..303S} presented the error bounds on the equivalent widths arising from uncertainties in the atomic data evaluated through Monte Carlo simulation. He obtained that the limiting accuracy of neon abundance determinations based on Ne\ione\ is $\sim \pm$0.10 dex adopting his currently available atomic data. For systematic uncertainties due to uncertainties in stellar parameters, we also refer to \citet{1999ApJ...519..303S}.

  \begin{deluxetable*}{ccccc|cccc|cccc}
  \tablecaption{NLTE and LTE Neon abundances in three B type stars with three different model atoms \label{Netest}}
  \tablewidth{0pt}
   \tablehead{ \colhead{$\lambda$}  & \colhead{NLTE} & \colhead{NLTE} & \colhead{NLTE} &\colhead{LTE}  & \colhead{NLTE} & \colhead{ NLTE} & \colhead{NLTE} & \colhead{LTE}  & \colhead{ NLTE}   & \colhead{NLTE} & \colhead{NLTE} & \colhead{ LTE}  \\
   \colhead{\AA} & \colhead{Model 1} & \colhead{Model 2}& \colhead{Model 3} &\colhead{} & \colhead{Model 1} & \colhead{Model 2}& \colhead{Model 3} &\colhead{} & \colhead{Model 1} & \colhead{Model 2}& \colhead{Model 3} 
  &\colhead{}     }
    \decimalcolnumbers
    \startdata
               & \multicolumn{4}c{HD~36959 (HR~1886) }  &  \multicolumn{4}c{HD~61068 (HR~2928) } &   \multicolumn{4}c{HD~36591 (HR~1861) } \\\hline
   {\bf  Ne\ione\ }   &       &        &      &       &       &        &        &       &        &       &        &        \\
  5852.            & \nodata&\nodata &\nodata &\nodata    &\nodata &\nodata  &\nodata   &\nodata     & 7.87  & 7.93   & 8.00   & 8.23  \\
  6096.            &  8.12  & 8.13   & 8.21   &  8.43     &8.05    & 8.07    & 8.14     &  8.36      & 8.14  & 8.17   & 8.24   & 8.44 \\              
  6143.            &  7.97  & 8.06   & 8.20   &  8.53     &7.94    & 8.00    & 8.12     &  8.47      & 7.99  & 8.07   & 8.19   & 8.55 \\             
  6266.            &\nodata &\nodata  &\nodata &\nodata   &\nodata & \nodata & \nodata  &\nodata     & 8.18  & 8.18   & 8.22   & 8.34  \\
  6334.            &  7.96  & 8.04   & 8.17   &  8.50     &8.03    & 8.10    & 8.23     &  8.58      & 7.96  & 8.03   & 8.15   & 8.50 \\             
  6382.            &  7.93  & 8.03   & 8.13   &  8.46     &8.02    & 8.12    & 8.22     &  8.56      & 7.97  & 8.03   & 8.19   & 8.51 \\             
  6402.            &  7.94  & 8.02   & 8.18   &  8.64     &7.92    & 7.99    & 8.13     &  8.57      & 7.98  & 8.05   & 8.18   & 8.63 \\             
  6506.            &  7.92  & 8.01   & 8.11   &  8.46     &7.96    & 8.04    & 8.15     &  8.51      & 7.93  & 8.01   & 8.12   & 8.47 \\
  6598.            &  8.04  & 8.09   & 8.17   &  8.44     &\nodata & \nodata & \nodata  &  \nodata   &\nodata&\nodata &\nodata &\nodata \\
  6717.            &  8.01  & 8.08   & 8.17   &  8.46     &\nodata & \nodata & \nodata  &  \nodata   &\nodata&\nodata &\nodata &\nodata \\
  7032.            &  7.99  & 8.08   & 8.23   &  8.58     &7.96    & 8.03    & 8.17     &  8.54      & 7.94  & 8.01   & 8.15   & 8.52  \\\hline
Mean Ne\ione\      &  7.99  & 8.06 & 8.17 & 8.50 & 7.98 & 8.05   & 8.17     & 8.51   & 8.00 & 8.05 & 8.16 & 8.47  \\
  $\sigma$         &  0.06  & 0.04   & 0.04   &  0.07     &0.05    & 0.05    & 0.04     &  0.08      & 0.10  & 0.08   & 0.07   & 0.12  \\\hline
 {\bf Ne\ii\  }    &       &        &      &       &       &        &        &       &        &       &        &        \\
  4150.            &\nodata &\nodata &\nodata & \nodata   &\nodata &\nodata  &\nodata   & \nodata    & 7.99  & 8.00   & 8.09   &  8.06 \\
  4219.            &  7.97  & 7.96   & 8.04   &  8.03     & 7.94   & 7.93    & 8.01     &  8.00      & 8.01  & 7.99   & 8.09   &  8.07 \\
  4220.            &\nodata &\nodata &\nodata & \nodata   &\nodata &\nodata  &\nodata   & \nodata    & 8.10  & 8.09   & 8.17   &  8.16 \\
  4231.            &\nodata &\nodata &\nodata & \nodata   &\nodata &\nodata  &\nodata   & \nodata    & 8.08  & 8.07   & 8.17   &  8.15 \\
  4239.            &  8.05  & 8.04   & 8.11   &  8.10     & 8.02   & 8.01    & 8.09     &  8.08      & 8.06  & 8.05   & 8.14   &  8.12 \\
  4250.            &\nodata &\nodata &\nodata & \nodata   &\nodata &\nodata  &\nodata   & \nodata    & 7.99  & 7.98   & 8.07   &  8.05 \\
  4391.            &  8.04  & 7.98   & 8.07   &  8.05     & 7.95   & 7.89    & 7.99     &  7.97      & 8.08  & 8.01   & 8.12   &  8.08 \\
  4412.            &\nodata &\nodata &\nodata & \nodata   &\nodata &\nodata  &\nodata   & \nodata    & 7.99  & 7.98   & 8.07   &  8.08 \\\hline
  Mean Ne\ii\      &  8.02  & 7.99 & 8.07 &  8.06     & 7.97   & 7.94 & 8.03    &  8.02      & 8.04  & 8.02   & 8.12   & 8.10  \\
  $\sigma$         &  0.04  & 0.04   & 0.04   &  0.04     & 0.04   & 0.06    & 0.05 &  0.06      & 0.05  & 0.04   & 0.04   & 0.04  \\ \hline
Ne\ione\ $-$Ne\ii\ &-0.03   & 0.07 & 0.10 & 0.44   & 0.01 & 0.11 &  0.14   & 0.49  & -0.04 & 0.03 & 0.04 & 0.37 \\ \hline
  \enddata
    \tablecomments{{\bf Note.} Model 1: the model atom as described in Sect. \ref{subSect:atom} with electron collisional data for Ne\ione\ from \citet{2012PhRvA..86b2717Z, 2012PhRvA..85f2710Z} for 461 transitions and collisional data for Ne\ii\ from \citet{2007JPhB...40.2969W} for 3078 transitions
 and formula of \citet{1962ApJ...136..906V} and $\Omega$=1
 for the remaining allowed and forbidden transitions. Model 2: model atom, where electron impact excitation  was included neither for Ne\ione\ nor for Ne\ii. 
 Model 3: model atom, where for allowed and forbidden transitions of the Ne\ione\ and Ne\ii\ the only formula of \citet{1962ApJ...136..906V} and $\Omega$=1 were employed. } 
  \end{deluxetable*}

   \subsection{Neon abundances in B-stars}

\begin{figure*}
 \begin{center}
 \includegraphics[scale=0.5]{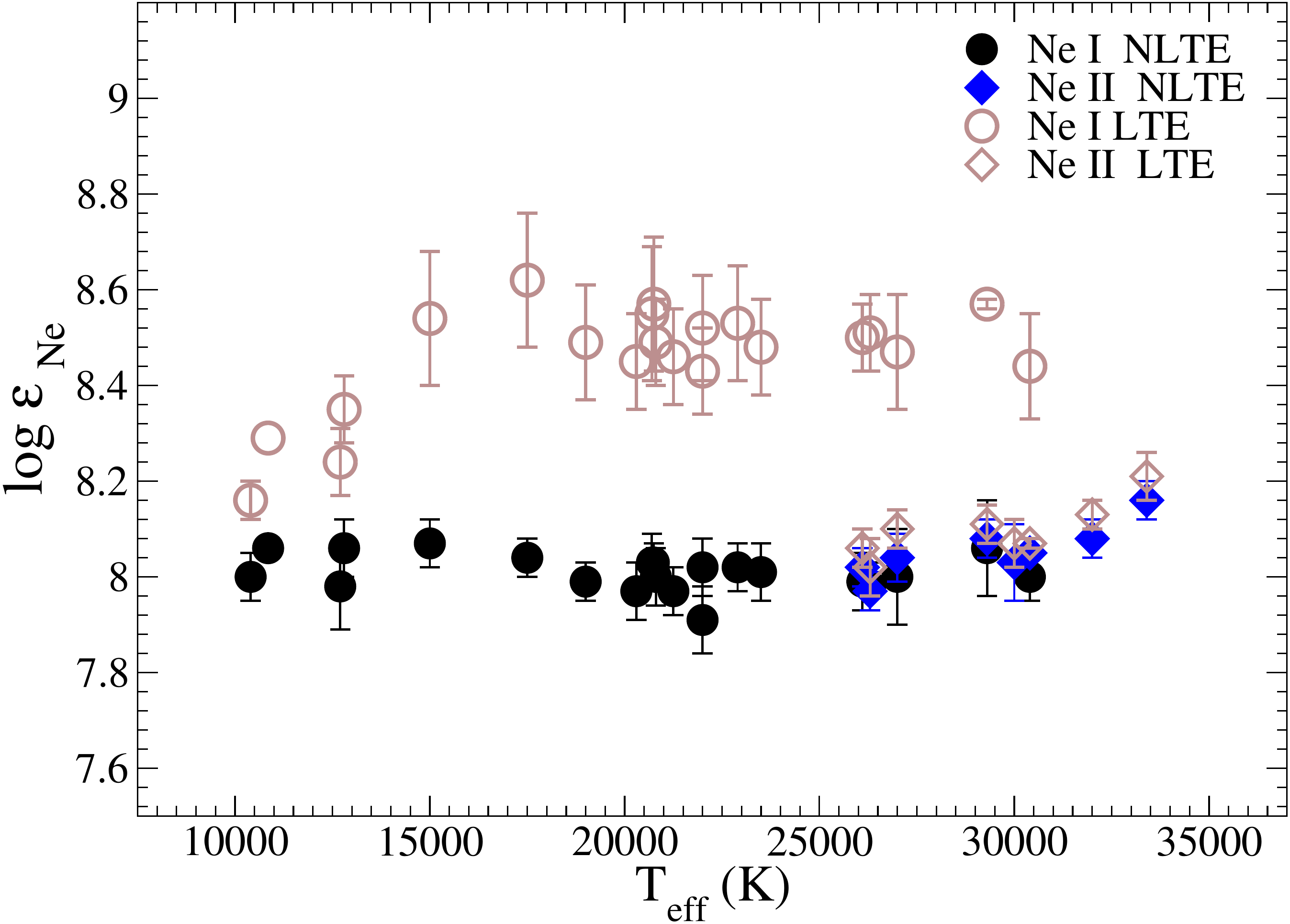}
 \caption{NLTE and LTE abundances from the Ne\ione\ and Ne\ii\ lines for the sample stars. The error bars correspond to the dispersion in the single-line measurements
about the mean.}
 \label{ABN}
 \end{center}
 \end{figure*}

  In this section, we derive the neon abundance of the sample stars with effective temperatures from 10\,250~K to 33\,400~K using lines of Ne\ione\ and Ne\ii\ in the visible and near-IR spectral ranges. 
  Everywhere, the Model 1 was used for the NLTE neon abundance determinations. The abundance results are presented in Table~\ref{tab_abn1} for individual lines.  Summarizing for NLTE neon abundances in the sample stars is presented in Table~\ref{tab_abn2}. The mean abundances from Ne\ione\ + Ne\ii\ lines (column 10) were calculated from all lines of both ionization stages.  
  Mean NLTE and LTE neon abundances with the standard deviations $\sigma$ for each star are shown in Fig.~\ref{ABN}.  
  For five stars, where lines of both ionization stages are available, the NLTE abundances derived from the Ne\ione\ and the Ne\ii\ lines agree within the error bars, while, in LTE, the abundance differences can reach up to 0.53~dex. 
  The NLTE corrections for Ne\ione\ lines are negative for all considered lines and can reach up to $-$1.10~dex. 
  The deviations from LTE for the most Ne\ii\ lines are small and do not exceed 0.11~dex in absolute value.

  Below we briefly comment on some individual lines. 
  
  {\bf Ne\ione\ 6402~\AA } is the strongest line and it is observed in the spectra of the stars in the wide temperature range of \Teff\ from 10\,250~K to 30\,400~K. The NLTE effect is strongest for this line compared to other lines.
    $\Delta_{\rm NLTE}$ for this line ranges from $-$0.24 dex (10\,400/3.50) to $-$1.10 dex (17\,500/3.8). 
  This line appears in relatively cool stars with \Teff\ = 10\,400~K, where other lines are not available yet, and it also appears in the hotter stars with \Teff\ = 30\,400~K, where other lines of Ne\ione\ are not visible any more. 
    
  {\bf Ne\ione\ 6143~\AA } line is possible to measure in the range of \Teff\ from 10\,400~K to 27\,000~K. The largest NLTE effect of this line is found in $\iota$~Her (17\,500/3.8) with $\Delta_{\rm NLTE}$ = $-$0.71 dex,
and the effect becomes to be smaller in cooler or hotter atmospheres. In  21~Peg (10\,400/3.8), $\Delta_{\rm NLTE}$ = $-$0.17 dex; and in HD~36591 (27\,000/4.12), $\Delta_{\rm NLTE}$ = $-$0.56 dex.

  Below we briefly comment on some representative stars.

  {\bf 21~Peg.} This is the coolest star of our sample, and the neon abundance, log~$\epsilon_{\rm Ne}$ = 8.00$\pm$0.05, was derived from the Ne\ione\ lines only.
   Hereafter, the statistical abundance error is the dispersion in the single line measurements about the mean:
  $\sigma = [\Sigma (x - x_i )^2 /(N - 1)]^{1/2}$, where N is the total number of lines used, x is their mean abundance, x$_i$ is the abundance derived from each individual line.
  
  {\bf HD~22136.} This star is a SPB-type variable \citep{2005AcA....55..375M}, and its line profiles are asymmetric. 
  The asymmetry causes uncertainties in the line fitting procedure, depending on the line intensity. 
  Nevertheless, the neon NLTE abundances based on 8 Ne\ione\ lines are consistent within 0.09~dex. 
  
  {\bf $\pi$~Cet.} Fifteen Ne\ione\ lines were measured in this star giving log~$\epsilon_{\rm Ne}$ = 8.06$\pm$0.06. 
  The NLTE effects for the Ne\ione\ lines are similar to those for HD~22136, although they are slightly stronger.    
 
\begin{figure*}
  \begin{minipage}{160mm}
  \parbox{0.5\linewidth}{\includegraphics[scale=0.29]{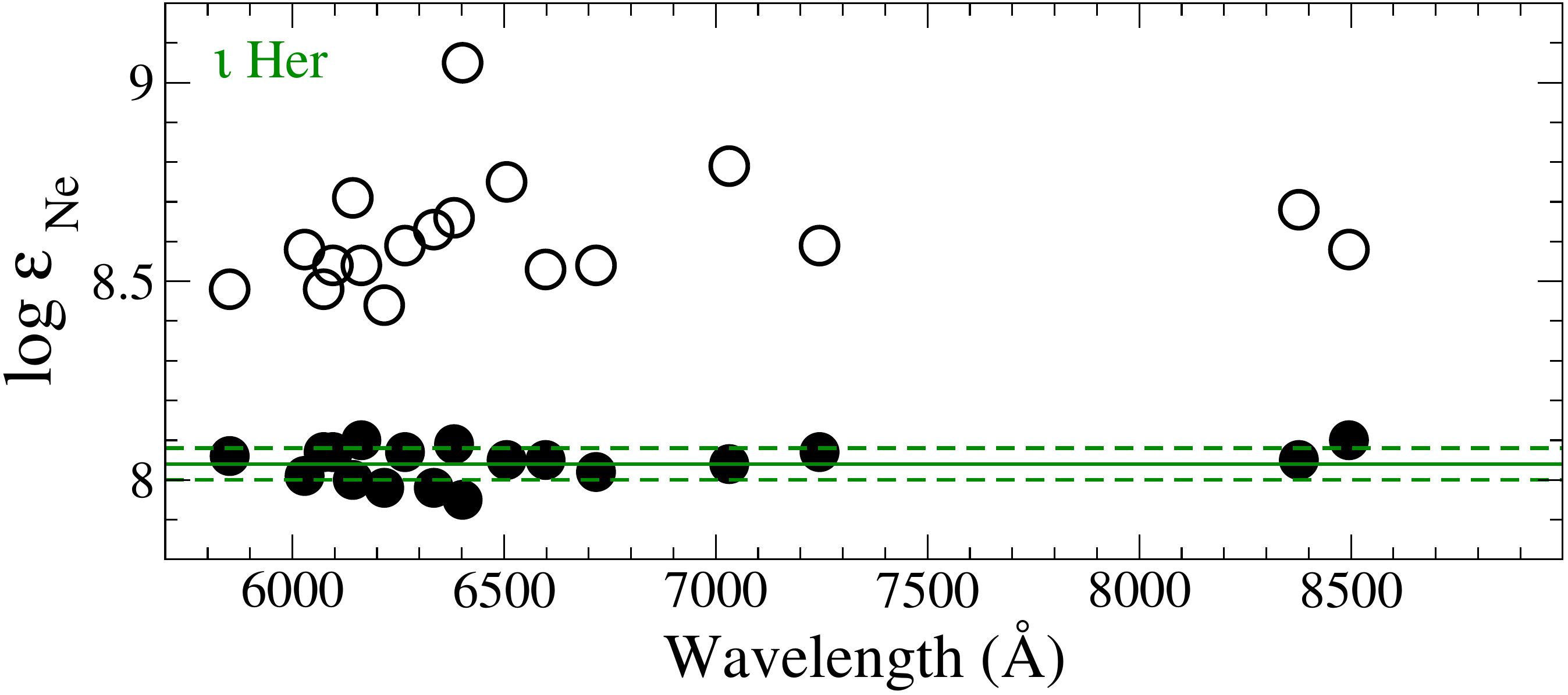}\\
  \centering}
  \parbox{0.5\linewidth}{\includegraphics[scale=0.29]{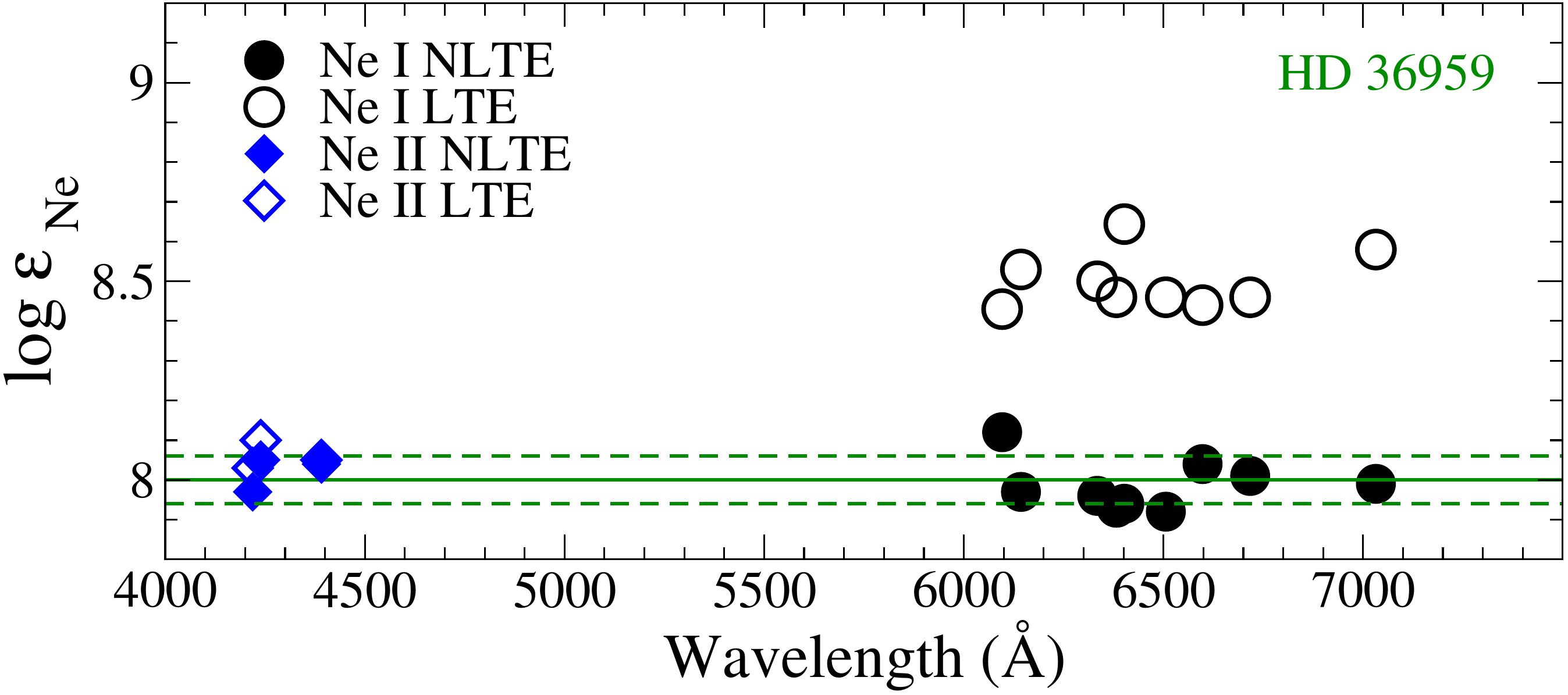}\\
  \centering}
  \hspace{1\linewidth}
  \hfill
  \\[0ex]
  \parbox{0.5\linewidth}{\includegraphics[scale=0.29]{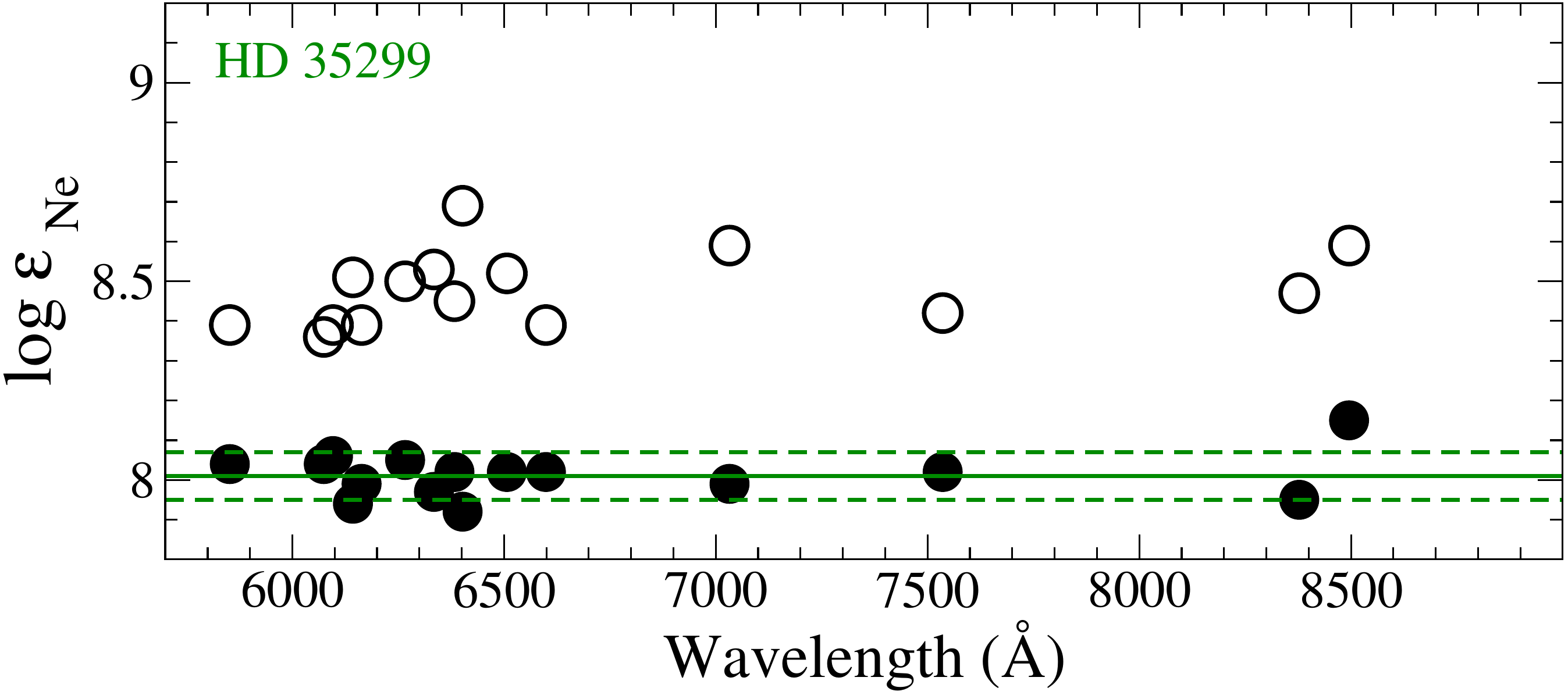}\\
  \centering}
  \parbox{0.5\linewidth}{\includegraphics[scale=0.29]{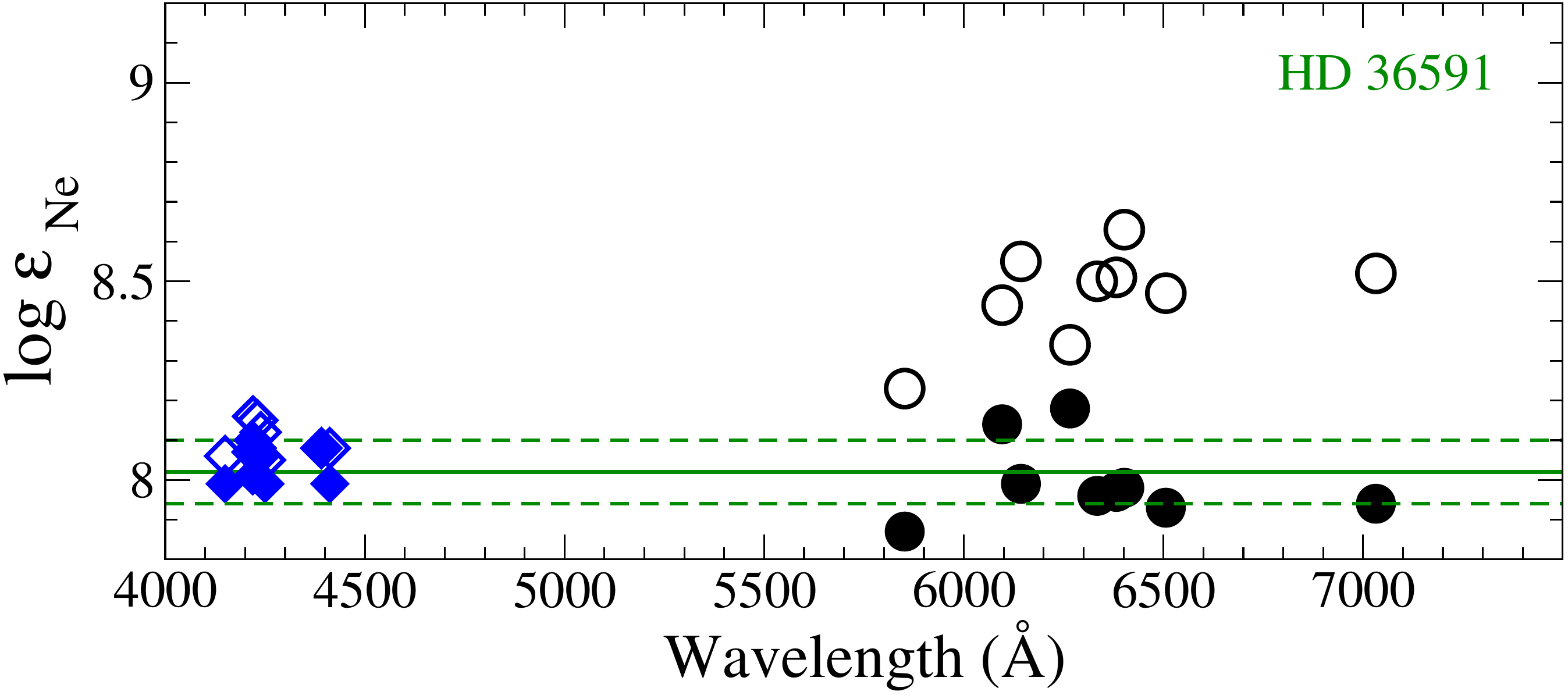}\\
  \centering}
  \hspace{1\linewidth}
  \hfill
  \\[0ex]
  \caption{Neon LTE (open symbols) and NLTE (filled symbols) abundances of a few program stars derived from lines of Ne\ione\ (circles) and Ne\ii\ (rhombi) in a wide spectral region. 
  In each panel, the solid line represents the NLTE abundance averaged over all Ne\ione\ and Ne\ii\ (where possible) lines. Dashed lines represent the standard deviation.
   }
  \label{pics}
  \end{minipage}
  \end{figure*}

  {\bf $\iota$~Her.} The largest number of Ne\ione\ lines are measured in this star. From eighteen Ne\ione\ lines we obtain log~$\epsilon_{\rm Ne}$ = 8.04$\pm$0.04 (Fig.~\ref{pics}). 
  The NLTE corrections are negative for all Ne\ione\ lines and they reach maximum in absolute value in the atmosphere of this star. 
  The smallest NLTE correction is $\Delta_{\rm NLTE}$ = $-$0.41 dex for the Ne\ione\ 6074~\AA\ line and the largest one is $\Delta_{\rm NLTE}$ = $-$1.10 dex for the Ne\ione\ 6402~\AA\ line.
  
  {\bf $\tau$~Sco.} This star (HD~149438) is the third hottest magnetic star known \citep{2006MNRAS.370..629D} and its magnetic structure is unusually complex for a hot star. 
  $\tau$~Sco is a member of the Upper Sco association, however, it is a very young star with age $<$ 2 Myr compared with the association age of 11 Myr \citep{2012ApJ...746..154P}. The unusual characteristics of $\tau$~Sco could point to a blue straggler nature, due to a binary merger \citep{2014A&A...566A...7N}. The blue straggler nature may explain the rejuvenating via a merger, the origin of its magnetic field according to the mechanism suggested by \citet{2009MNRAS.400L..71F}, and the slow rotation of this star is explained as a result of angular-momentum loss. The merger scenario for $\tau$~Sco has been elaborated in detail recently by \citet{2019Natur.574..211S}.
  Due to its slow rotation, plenty of Ne\ii\ lines can be found in its spectrum. From six Ne\ii\ lines we got log~$\epsilon_{\rm Ne}$ = 8.08$\pm$0.04.
  
   {\bf $\upsilon$~Ori.} This star (HD~36512) is the hottest star in our sample and it is an outlier (Fig. \ref{ABN}). The neon abundance was derived from three Ne\ii\ lines only, giving log~$\epsilon_{\rm Ne}$ = 8.16$\pm$0.04.  This value is by 0.14~dex larger compared to homogeneous neon abundance for the entire sample of stars, for which the mean abundance is 8.02$\pm$0.04.  
  The NLTE corrections are $-$0.05, $-$0.02, and $-$0.08 for Ne\ii\ 4219, 4231, and 4391~\AA\ lines, respectively. 
  HD~36512 is a $\beta$~Cephei-type pulsator \citep{2005ApJS..158..193S}, and its line profiles are asymmetric. Nevertheless, the obtained neon abundances are consistent within 0.04~dex for three lines. 

  Nine stars of our sample (HD~34816, HD~35299, HD~35912, HD~36512, HD~36591, HD~36629, HD~36822, HD~36959, HD~37042) are the B main-sequence star members of the Orion OB1 (Ori OB1) association. 
  Orion molecular cloud and the Ori OB1 association are one of the most massive active star-forming regions within the 1 kpc centered on the Sun.
  Our derived neon abundances for these stars in this association are found to be homogeneous with averaged log~$\epsilon_{\rm Ne}$ = 8.03$\pm$0.06.

 \begin{deluxetable*}{cccccccccccccccccccccc}
 \tabletypesize{\scriptsize}
 \tablecaption{NLTE and LTE Neon abundances from Ne\ione\ and Ne\ii\ lines in 21 B-type star. \label{tab_abn1}}
\def\arraystretch{0.7}
\setlength{\tabcolsep}{2pt} 
\tabletypesize{\scriptsize }
\tablewidth{20pt}
\tablehead{
\colhead{$\lambda$ (\AA\,)}  & \colhead{\tiny NLTE} & \colhead{ \tiny LTE} & \colhead{ \tiny NLTE} & \colhead{ \tiny LTE}  & \colhead{ \tiny NLTE} & \colhead{ \tiny LTE}  & \colhead{ \tiny NLTE} & \colhead{ \tiny LTE} & \colhead{\tiny NLTE} & \colhead{\tiny LTE} & \colhead{\tiny NLTE} & \colhead{ \tiny LTE}  & \colhead{ \tiny NLTE} & \colhead{ \tiny LTE}
}
\startdata
 \multicolumn{15}c{ }      \\
 &\multicolumn2c{21~Peg}&\multicolumn2c{134~Tau} &\multicolumn2c{HD~22136}  &\multicolumn2c{$\pi$ Cet} &\multicolumn2c{$\tau$~Her} &\multicolumn2c{$\iota$~Her}  &\multicolumn2c{HR 1820} \\ 
  \multicolumn{15}c{ }      \\\hline
 Ne\ione\ & \multicolumn{14}c{ }      \\ 
 5852.  & \nodata&\nodata &\nodata &\nodata  & 8.06    &  8.24  &  8.04 & 8.25   &\nodata&\nodata  & 8.06   & 8.48    & 7.97   & 8.34    \\
 6029.  & \nodata&\nodata &\nodata &\nodata  & \nodata &\nodata &  8.04 & 8.23   &\nodata&\nodata  & 8.01   & 8.58    & 7.99   & 8.34    \\
 6074.  & \nodata&\nodata &\nodata &\nodata  & \nodata &\nodata &  8.09 & 8.30   & 8.13  &  8.45   & 8.07   & 8.48    & 8.03   & 8.42    \\
 6096.  & 7.95   & 8.08   &\nodata &\nodata  & 7.99    & 8.17   &  8.09 & 8.32   & 8.07  &  8.43   & 8.07   & 8.54    & 8.01   & 8.40    \\
 6143.  & 8.01   & 8.18   &\nodata &\nodata  & 7.94    & 8.23   &  8.00 & 8.36   & 8.05  &  8.62   & 8.00   & 8.71    & 7.95   & 8.56    \\
 6163.  &\nodata &\nodata &\nodata &\nodata  & \nodata &\nodata &  8.07 & 8.30   & 8.07  &  8.39   & 8.10   & 8.54    & 8.06   & 8.46    \\
 6217.  &\nodata &\nodata &\nodata &\nodata  & \nodata &\nodata &  8.00 & 8.25   & 8.06  &  8.40   & 7.98   & 8.44    &\nodata &\nodata  \\
 6266.  &\nodata &\nodata &\nodata &\nodata  & \nodata &\nodata &  8.10 & 8.38   & 8.16  &  8.58   & 8.07   & 8.59    & 8.04   & 8.51    \\
 6334.  & 7.97   & 8.15   &\nodata &\nodata  & 7.92    & 8.20   &  8.03 & 8.39   & 8.07  &  8.58   & 7.98   & 8.63    & 7.96   & 8.54    \\
 6382.  & 8.08   & 8.18   &\nodata &\nodata  & 8.07    & 8.29   &  8.06 & 8.33   &\nodata&\nodata  & 8.09   & 8.66    & 8.01   & 8.50    \\
 6402.  & 7.97   & 8.21   &  8.06  & 8.29    & 7.81    & 8.18   &  7.95 & 8.46   & 7.99  &  8.85   & 7.95   & 9.05    & 7.88   & 8.78    \\
 6506.  & 8.03   & 8.18   &\nodata &\nodata  & 8.09    & 8.37   &  8.12 & 8.44   & 8.05  &  8.60   & 8.05   & 8.75    & 7.98   & 8.56    \\
 6598.  &\nodata &\nodata &\nodata &\nodata  & \nodata &\nodata &  8.15 & 8.40   &\nodata&\nodata  & 8.05   & 8.53    & 8.01   & 8.45    \\
 6717.  &\nodata &\nodata &\nodata &\nodata  & \nodata &\nodata &\nodata&\nodata & 8.05  &  8.46   & 8.02   & 8.54    & 7.99   & 8.45    \\
 7032.  & 7.98   & 8.14   &\nodata &\nodata  & 7.97    & 8.27   &  8.05 & 8.42   &\nodata&\nodata  & 8.04   & 8.79    & 7.96   & 8.61    \\
 7245.  &\nodata &\nodata &\nodata &\nodata  & \nodata &\nodata &\nodata&\nodata &\nodata&\nodata  & 8.07   & 8.59    &\nodata &\nodata  \\
 7535.  &\nodata &\nodata &\nodata &\nodata  & \nodata &\nodata &  8.16 & 8.35   &\nodata&\nodata  &\nodata &\nodata  & 7.97   & 8.36    \\ 
 8377.  &\nodata &\nodata &\nodata &\nodata  & \nodata &\nodata &\nodata&\nodata &\nodata&\nodata  & 8.05   & 8.68    & 8.03   & 8.59    \\
 8495.  &\nodata &\nodata &\nodata &\nodata  & \nodata &\nodata &\nodata&\nodata &\nodata&\nodata  & 8.10   & 8.58    &\nodata &\nodata  \\\hline
 Mean   & 8.00   & 8.16   & 8.06   & 8.29    & 7.98  & 8.24&  8.06& 8.35& 8.07  & 8.54 & 8.04   &  8.62  & 7.99   & 8.49    \\ 
$\sigma$& 0.05  & 0.04   &\nodata &\nodata&  0.09   & 0.07   &  0.06 & 0.07   & 0.05  & 0.14    & 0.04   & 0.14    & 0.04   & 0.12    \\ \hline
 \multicolumn{15}c{ }      \\
 &\multicolumn2c{HIP 26000} &\multicolumn2c{ $o$ Tau} &\multicolumn2c{$\zeta$ Cas} &\multicolumn2c{$\chi$ Cen} &\multicolumn2c{$\delta$ Cet } &\multicolumn2c{$\nu$ Eri } &\multicolumn2c{$\gamma$ Peg}   \\
\multicolumn{15}c{ }      \\\hline
 Ne\ione\ & \multicolumn{14}c{ }      \\
5852.   & 7.99 & 8.36    & 8.04  & 8.42     & 8.01  & 8.40      & 7.98  & 8.34    & 7.99  & 8.37   &\nodata&\nodata & 8.01  & 8.38      \\
6029.   & 8.04 & 8.40    &\nodata&\nodata   &\nodata&\nodata    &\nodata&\nodata  & 8.05  & 8.43   &\nodata&\nodata &\nodata&\nodata    \\
6074.   & 8.00 & 8.33    & 8.12  & 8.48     & 8.11  & 8.48      & 8.08  & 8.43    & 8.03  & 8.39   &\nodata&\nodata & 8.08  & 8.42      \\
6096.   & 8.01 & 8.39    &\nodata&\nodata   & 8.06  & 8.46      & 8.07  & 8.43    & 7.97  & 8.34   &\nodata&\nodata & 8.05  & 8.41      \\
6143.   & 7.88 & 8.45    & 7.99  & 8.60     & 7.98  & 8.61      & 7.95  & 8.53    & 7.91  & 8.48   & 7.89  & 8.42   & 7.99  & 8.58      \\
6163.   & 8.03 & 8.42    & 8.07  & 8.47     & 8.04  & 8.51      & 8.05  & 8.44    & 7.94  & 8.36   &\nodata&\nodata & 8.03  & 8.45      \\
6266.   & 7.98 & 8.43    &\nodata&\nodata   &\nodata&\nodata    &\nodata&\nodata  & 7.92  & 8.40   &\nodata&\nodata & 7.99  & 8.46      \\
6334.   & 7.89 & 8.45    & 8.00  & 8.59     & 8.01  & 8.62      & 7.97  & 8.52    & 7.92  & 8.49   &\nodata&\nodata & 7.94  & 8.51      \\
6382.   & 8.00 & 8.47    & 8.08  & 8.56     & 8.04  & 8.52      & 8.02  & 8.47    & 7.97  & 8.43   & 8.00  & 8.42   & 8.06  & 8.52      \\
6402.   & 7.85 & 8.69    & 7.95  & 8.83     & 7.96  & 8.89      & 7.89  & 8.69    & 7.88  & 8.72   & 7.84  & 8.48   & 7.93  & 8.76       \\
6506.   & 7.94 & 8.48    & 8.04  & 8.60     & 8.00  & 8.56      & 8.02  & 8.55    & 7.94  & 8.47   & 7.85  & 8.29   & 8.00  & 8.53      \\                                                                                                    
6598.   & 8.03 & 8.45    & 8.07  & 8.51     & 8.06  & 8.49      & 8.06  & 8.47    & 8.05  & 8.48   &\nodata&\nodata &\nodata&\nodata    \\
7032.   & 7.94 & 8.56    & 8.04  & 8.69     & 8.04  & 8.71      & 7.96  & 8.57    & 7.97  & 8.61   & 7.96  & 8.54   & 8.01  & 8.63     \\
7535.   & 7.90 & 8.28    & 7.89  & 8.27     &\nodata&\nodata    &\nodata&\nodata  & 7.96  & 8.37   &\nodata&\nodata & 7.99  & 8.40     \\
8377.   & 7.99 & 8.55    & 8.01  & 8.56     &\nodata&\nodata    &\nodata&\nodata  & 7.95  & 8.51   &\nodata&\nodata & 8.02  & 8.59     \\
8495.   & 8.06 & 8.53    &\nodata&\nodata   &\nodata&\nodata    & 8.00  & 8.41    & 8.02  & 8.47   &\nodata&\nodata & 8.15  & 8.63     \\\hline
Mean &7.97 &8.45& 8.03 &8.55 &8.03   &8.57  & 8.00 & 8.49 & 7.97 & 8.46 & 7.91  & 8.43 & 8.02  & 8.52    \\
$\sigma$& 0.06 & 0.10    & 0.06  & 0.14     &0.04   & 0.14      & 0.06 & 0.09    & 0.05 & 0.10    & 0.07  & 0.09   &  0.06  & 0.11     \\\hline
\multicolumn{15}c{ }      \\
  &\multicolumn2c{ $\alpha$ Pix}  &\multicolumn2c{ HR 1781} &\multicolumn2c{$\theta^2$ Ori B} &\multicolumn2c{$\phi^1$ Ori} &\multicolumn2c{$\lambda$ Lep } &\multicolumn2c{ $\tau$ Sco } &\multicolumn2c{ $\upsilon$ Ori}  \\
\multicolumn{15}c{ }      \\\hline
 Ne\ione\  & \multicolumn{14}c{ }      \\
  5852.    & 7.94  & 8.30     & 8.04   & 8.39     &\nodata &\nodata   &\nodata &\nodata  &\nodata  &\nodata&\nodata &\nodata  &\nodata &\nodata   \\
  6074.    & 8.09  & 8.41     & 8.04   & 8.36     &\nodata &\nodata   &\nodata &\nodata  &\nodata  &\nodata&\nodata &\nodata  &\nodata &\nodata   \\
  6096.    &\nodata&\nodata   & 8.06   & 8.39     &\nodata &\nodata   &\nodata &\nodata  &\nodata  &\nodata&\nodata &\nodata  &\nodata &\nodata   \\
  6143.    & 7.99  & 8.56     & 7.94   & 8.51     &\nodata &\nodata   &\nodata &\nodata  &\nodata  &\nodata&\nodata &\nodata  &\nodata &\nodata   \\
  6163.    & 8.06  & 8.52     & 7.99   & 8.39     &\nodata &\nodata   &\nodata &\nodata  &\nodata  &\nodata&\nodata &\nodata  &\nodata &\nodata   \\
  6266.    & 8.06  & 8.55     & 8.05   & 8.50     &\nodata &\nodata   &\nodata &\nodata  &\nodata  &\nodata&\nodata &\nodata  &\nodata &\nodata   \\
  6334.    & 8.00  & 8.58     & 7.97   & 8.53     &\nodata &\nodata   &\nodata &\nodata  &\nodata  &\nodata&\nodata &\nodata  &\nodata &\nodata   \\
  6382.    & 8.04  & 8.48     & 8.02   & 8.45     &\nodata &\nodata   &\nodata &\nodata  &\nodata  &\nodata&\nodata &\nodata  &\nodata &\nodata   \\
  6402.    & 7.96  & 8.68     & 7.92   & 8.69     & 7.99   & 8.57     &\nodata &\nodata  & 7.96    &8.52   &\nodata &\nodata  &\nodata &\nodata   \\
  6506.    & 8.02  & 8.50     & 8.02   & 8.52     & 8.13   & 8.56     &\nodata &\nodata  & 8.03    &8.36   &\nodata &\nodata  &\nodata &\nodata   \\
  6598.    &\nodata&\nodata   & 8.02   & 8.39     &\nodata &\nodata   &\nodata &\nodata  &\nodata  &\nodata&\nodata &\nodata  &\nodata &\nodata   \\
  7032.    & 8.07  & 8.69     & 7.99   & 8.59     &\nodata &\nodata   &\nodata &\nodata  &\nodata  &\nodata&\nodata &\nodata  &\nodata &\nodata   \\ 
  7535.    &\nodata&\nodata   & 8.02   & 8.42     &\nodata &\nodata   &\nodata &\nodata  &\nodata  &\nodata&\nodata &\nodata  &\nodata &\nodata   \\
  8377.    &\nodata&\nodata   & 7.95   & 8.47     &\nodata &\nodata   &\nodata &\nodata  &\nodata  &\nodata&\nodata &\nodata  &\nodata &\nodata   \\
  8495.    &\nodata&\nodata   & 8.15   & 8.59     &\nodata &\nodata   &\nodata &\nodata  &\nodata  &\nodata&\nodata &\nodata  &\nodata &\nodata   \\\hline                                                                                                          
  Mean     & 8.02 & 8.53 & 8.01 & 8.48   & 8.06 & 8.57    &\nodata &\nodata  & 8.00    & 8.44 &\nodata &\nodata  &\nodata &\nodata   \\ 
  $\sigma$ & 0.05  & 0.12     & 0.06   & 0.10 & 0.10   & 0.01     &\nodata &\nodata  & 0.05    & 0.11  &\nodata &\nodata  &\nodata &\nodata   \\ \hline
 Ne\ii\ & \multicolumn{14}c{ }      \\                                                                          
4150.      &\nodata &\nodata &\nodata &\nodata   &\nodata &\nodata   &\nodata &\nodata  &\nodata  &\nodata & 8.03  & 8.09    &\nodata &\nodata   \\ 
4219.      &\nodata &\nodata &\nodata &\nodata   & 8.04   & 8.07     & 8.07   & 8.10    & 8.05    & 8.07   & 8.06  & 8.13    &   8.12  & 8.17     \\ 
4220.      &\nodata &\nodata &\nodata &\nodata   &\nodata &\nodata   & 8.08   & 8.11    & 8.05    & 8.07   & 8.08  & 8.10    &\nodata  &\nodata   \\ 
4231.      &\nodata &\nodata &\nodata &\nodata   & 8.09   & 8.15     & 7.97   & 8.02    & 8.04    & 8.08   & 8.12  & 8.18    &   8.18  & 8.20     \\ 
4239.      &\nodata &\nodata &\nodata &\nodata   &\nodata &\nodata   & 7.91   & 8.00    &\nodata  &\nodata & 8.08  & 8.14    &\nodata  &\nodata   \\ 
4391.      &\nodata &\nodata &\nodata &\nodata   & 8.12   & 8.10     & 8.11   & 8.10    &\nodata  &\nodata & \nodata&\nodata &   8.18  & 8.26     \\
4412.      &\nodata &\nodata &\nodata &\nodata   &\nodata &\nodata   &\nodata &\nodata  &\nodata  &\nodata & 8.13  & 8.15    &\nodata  &\nodata   \\  \hline
 Mean      &\nodata &\nodata &\nodata &\nodata   & 8.08   & 8.11     & 8.03   & 8.07    & 8.05    & 8.07   & 8.08  & 8.13 &   8.16  & 8.21     \\ 
$\sigma$   &\nodata &\nodata &\nodata &\nodata   & 0.04   & 0.04     & 0.08   & 0.05    & 0.01    & 0.01   & 0.04  & 0.03    &   0.04  & 0.05     \\ \hline
\enddata                                                                                                                                                                  
\end{deluxetable*}

   \begin{deluxetable*}{cllccclcccc}
    \tablecaption{NLTE Neon abundances in the sample stars \label{tab_abn2}}
    \tablewidth{0pt}
    \tablehead{  
    \colhead{Number} &\colhead{Star} & \colhead{Name} & \colhead{Ne\ione}& \colhead{$\sigma$} & \colhead{N}& \colhead{Ne\ii }& \colhead{$\sigma$} & \colhead{N} & \colhead{Ne\ione+Ne\ii} & \colhead{$\sigma$}   }
    \decimalcolnumbers
    \startdata
    1   & HD~209459 & 21~Peg              &  8.00 & 0.05 & 7  &\nodata &\nodata & --   & 8.00  &  0.05  \\         
    2   & HD~38899  & 134~Tau             &  8.06 &\nodata & 1   &\nodata &\nodata & --   & 8.06  & \nodata     \\         
    3   & HD~22136  &                     & 7.98 & 0.09   & 8   &\nodata &\nodata & --   & 7.98 &  0.09      \\         
    4   & HD~17081  &  $\pi$ Cet          & 8.06 & 0.06   & 15  &\nodata &\nodata & --   & 8.06 &  0.06       \\        
    5   & HD~147394 & $\tau$~Her          &  8.07 & 0.05 & 10  &\nodata &\nodata & -- & 8.07  & 0.05     \\         
    6   & HD~160762 & $\iota$~Her         &  8.04 & 0.04   & 18  &\nodata &\nodata & --   & 8.04  &  0.04      \\         
    7   & HD~35912  & HR~1820             &  7.99 & 0.04   & 16  &\nodata &\nodata & --   & 7.99  &  0.04      \\         
    8   & HD~36629  & HIP~26000           &  7.97 & 0.06   & 16  &\nodata &\nodata & --   & 7.97&  0.06      \\         
    9   & HD~35708  & $o$~Tau             &  8.03 & 0.06   & 12  &\nodata &\nodata & --   & 8.03  &  0.06       \\        
    10  & HD~3360   & $\zeta$ Cas         &  8.03 & 0.04   & 11  &\nodata &\nodata & --   & 8.03  &  0.04      \\         
    11  & HD~122980 & $\chi$ Cen          &  8.00 & 0.06 & 12  &\nodata &\nodata & --   &  8.00&  0.06       \\        
    12  & HD~16582  & $\delta$ Cet        &  7.97 & 0.05   & 16  &\nodata &\nodata & --   & 7.97  &  0.05       \\        
    13  & HD~29248  & $\nu$ Eri           &  7.91 & 0.07   & 5   &\nodata &\nodata & --   & 7.91  &  0.07      \\        
    14  & HD~886    & $\gamma$ Peg        &  8.02 & 0.06 & 14  &\nodata &\nodata & --   & 8.02  & 0.06     \\         
    15  & HD~74575  & $\alpha$ Pix        & 8.02 & 0.05   & 10  &\nodata &\nodata & --   & 8.02 &  0.05      \\         
    16  & HD~35299  & HR~1781             & 8.01 & 0.06   & 15  &\nodata &\nodata & --   & 8.02  &  0.06      \\  
    17  & HD~36959  & HR~1886             &  7.99 & 0.06   & 9   &   8.02 &   0.04 & 3    & 8.00  &  0.06      \\         
    18  & HD~61068  & HR~2928        &  7.98 & 0.05   & 7   &   7.97 &   0.04 & 3    & 7.98  &  0.04 \\         
    19  & HD~36591  & HR~1861        &  8.00 & 0.10   & 9   &   8.04 &   0.05 & 8    & 8.02  &  0.08      \\  
    20  & HD~37042  & $\theta^2$ Ori B   & 8.06 & 0.10  & 2   &   8.08 &   0.04 & 3    & 8.07  &  0.06  \\ 
    21  & HD~36822  & $\phi^1$ Ori       &\nodata&\nodata & --  &   8.03 &   0.08 & 5    & 8.03  &  0.08       \\ 
    22  & HD~34816  & $\lambda$ Lep      &  8.00 & 0.05   & 2   &   8.05 &   0.01 & 3    & 8.03  & 0.05  \\    
    23  & HD~149438 & $\tau$ Sco         &\nodata&\nodata & --  &   8.08 &   0.04 & 6    & 8.08  &  0.04       \\    
    24  & HD~36512  & $\upsilon$ Ori     &\nodata&\nodata & --  &   8.16 &   0.04 & 3    & 8.16  & 0.04       \\\hline  
      \multicolumn3c{Mean}                & 8.01 &        &     &  8.05 &        &   & 8.02 &       \\
      \multicolumn3c{$\sigma$}            &  0.04 &        &     &   0.06 &        &      & 0.05  &           \\\hline
    \enddata                                                
    \end{deluxetable*}


\subsection{Comparison with previous studies of B-type stars}\label{sect:Comp}

   Table~\ref{tabcomp}  presents comparisons of the Neon abundances obtained in the present study with previous
studies.
  Our NLTE abundance analyses of the Ne\ione\ and Ne\ii\ lines in 24 B-type stars indicates a mean neon abundance log~$\epsilon_{\rm Ne}$ = 8.02$\pm$0.05, which is in line with other studies from the literature.
  
  Our mean neon abundance is in good agreement with the results from \citet{2012AA...539A.143N}, who established a present-day cosmic abundance standard 
  from a sample of 29 early B-type stars and got neon abundance log~$\epsilon_{\rm Ne}$ = 8.09$\pm$0.05. 
  They used the DETAIL code and model atom of Ne\ione/Ne\ii\ by \citet{2008A&A...487..307M} that was updated with gf-values for Ne\ione\ lines from \citet{2004ADNDT..87....1F}.  
  
   Our calculations are in good agreements with the remaining results from earlier studies.
  \citet{2010PASJ...62.1239T} carried out neon abundance determinations for 64 mid- through late-B stars using
two Ne\ione\ lines at 6143~\AA\ and at 6163~\AA, taking the NLTE effects into account. Their model atom includes 94 terms of Ne\ione\ and 1034 radiative transitions taken from \citet{1995all..book.....K} atomic data. 
  The calculated collisional rates are largely based on a treatment by \citet{1973ApJ...184..151A}.  They
  derived log~$\epsilon_{\rm Ne}$ = 8.02$\pm$0.09.
  
  \citet{2008A&A...487..307M} presented a homogeneous NLTE abundance study of the optical Ne\ione\ and Ne\ii\ lines in a sample of 18 nearby, early B-type stars.
  Their analysis gives log~$\epsilon_{\rm Ne}$ = 7.97$\pm$0.07 that is by 0.05 dex lower than our result.
  They developed an extensive model atom consisted of 153, 78 and 5 levels for Ne\ione\, Ne\ii\ and Ne\iii, respectively. 
  Oscillator strengths for Ne\ione\ and Ne\ii\ were adopted from Breit-Pauli R-matrix (BPRM) calculations \citep{1993A&A...279..298H}.
  Collisional data for the levels up to 5f states of Ne\ione\ were taken from BPRM calculation as described by \citep{2008A&G....49f..23B}, and for Ne\ii\ -- from Intermediate-Coupling Frame-Transformation (ICFT) R-matrix calculations by \citet{2001JPhB...34.4401G}.
  
  \citet{2006ApJ...647L.143C} obtained NLTE neon abundances for a sample of 11 B-type stellar members of the Orion association and found an average log~$\epsilon_{\rm Ne}$ = 8.11$\pm$0.04 using an extensive TLUSTY model atom and NLTE model atmospheres \citep{1995ApJ...439..875H}. 
  The constructed Ne model atom consists of 79 levels of Ne\ione, 138 levels of Ne\ii, 38 levels of Ne\iii, 12 levels of Ne\iv, and ground state of Ne\v. The collisional excitation rates were considered using the formula of \citet{1962ApJ...136..906V}, including a modification for the neutral atoms proposed by \citet{1973ApJ...184..151A}.

  \citet{2003A&A...408.1065H} presented NLTE neon abundance studies of 9 bright B5--B9 stars, and obtained log~$\epsilon_{\rm Ne}$ = 8.16$\pm$0.14 dex using the Ne\ione\ 6402 and 6506~\AA\ lines
  and model atom consisted of 45 Ne\ione\ levels, 47 Ne\ii\ levels, and 120 transitions, for which the oscillator strengths were adopted from \citet{1998JPhB...31.5315S} and \citet{1975SAOSR.362.....K}.
  
  \citet{2000MNRAS.318.1264D} obtained log~$\epsilon_{\rm Ne}$ = 8.10$\pm$0.09 dex for 7 early A and late B stars based on one strong Ne\ione\ line at 6402~\AA\ using 31 level Ne\ione\ model atom and TLUSTY NLTE model atmospheres. 
 
  \citet{1999ApJ...519..303S} performed the NLTE calculations for Ne\ione\ 6402 and 6506~\AA\, lines. The new model atom developed by him consisted of 37 Ne\ione\ and 11 Ne\ii\ levels. Sigut adopted R-matrix calculations by \citet{1985JPhB...18.2967T} and \citet{1997JPhB...30.4609Z} for electron impact excitation of the 2p$^5$3s and 2p$^5$3p configurations from the ground state and for the collisional excitation of a few transitions between the 2p$^5$3s and 2p$^5$3p configurations, and employed the impact parameter approximation by \citet{Seaton1962} for the remaining allowed transitions. 
  \citet{1999ApJ...519..303S} recomputed the NLTE abundances in the atmospheres of 14 early B-type stars from the sample of \citet{1992ApJ...387..673G} 
   based on their own equivalent widths for the Ne\ione\ 6506~\AA\ line and adopted their atmospheric parameters. \citet{1999ApJ...519..303S} does not present an average neon abundance for the sample stars he investigated.
   We extracted the data for 15 stars from Fig.~7 of \citet{1999ApJ...519..303S} and obtained the mean value log~$\epsilon_{\rm Ne}$ = 8.14$\pm$0.24~dex. The departure coefficients computed by \citet{1999ApJ...519..303S} (Figure 2 in his paper) are very similar to ours (Figure \ref{DC17}, left panel). 
   
   \citet{1994A&A...282..867K} derived neon abundance of 12 nearby, early B-type stars using a set of Ne\ii\ lines and found a mean value: log(Ne) = 8.10$\pm$0.06 dex.  It was an LTE analysis, but taking into account that 
   NLTE effects for Ne\ii\ lines are quite small, the result can be considered for comparison as well.

   \begin{deluxetable*}{lccc}
   \tablecaption{Comparison of the Ne NLTE Abundances Derived in This Paper with the Literature \label{tabcomp}}
   \def\arraystretch{1.0}
   \setlength{\tabcolsep}{6pt} 
   \tablewidth{35pt}
   \tablehead{
   \colhead{Source} & \multicolumn{2}c{ Result} & \colhead{Number of Stars}     \\
   \colhead{ }     & \colhead{log~$\epsilon_{\rm Ne}$ }  & \colhead{$\sigma$}  & \colhead{}   
   }
   \startdata
    This Work                   & 8.02 & 0.05   &  24             \\
   Nieva $\&$ Przybilla (2012)  & 8.09 & 0.05   &  29             \\
   Takeda et al. (2010)         & 8.02 & 0.09   &  64             \\
   Morel $\&$ Butler (2008)     & 7.97 & 0.07   &  18             \\
   Cunha et al. (2006)          & 8.11 & 0.04   &  11             \\
   Hempel $\&$ Holweger (2003)  & 8.16 & 0.14   &  9              \\
   Dworetsky $\&$ Budaj (2000)  & 8.10 & 0.09   &  7              \\
   Sigut (1999)                 & 8.14 & 0.24   &  15             \\
   Kilian (1994--LTE))          & 8.10 & 0.06   &  12             \\ 
   \enddata                                                                                                                                                                  
   \end{deluxetable*}

 \subsection{Comparison with the Sun and other objects in the Galaxy}\label{sect:Comp2}

\begin{figure*}
 \begin{center}
 \includegraphics[scale=0.5]{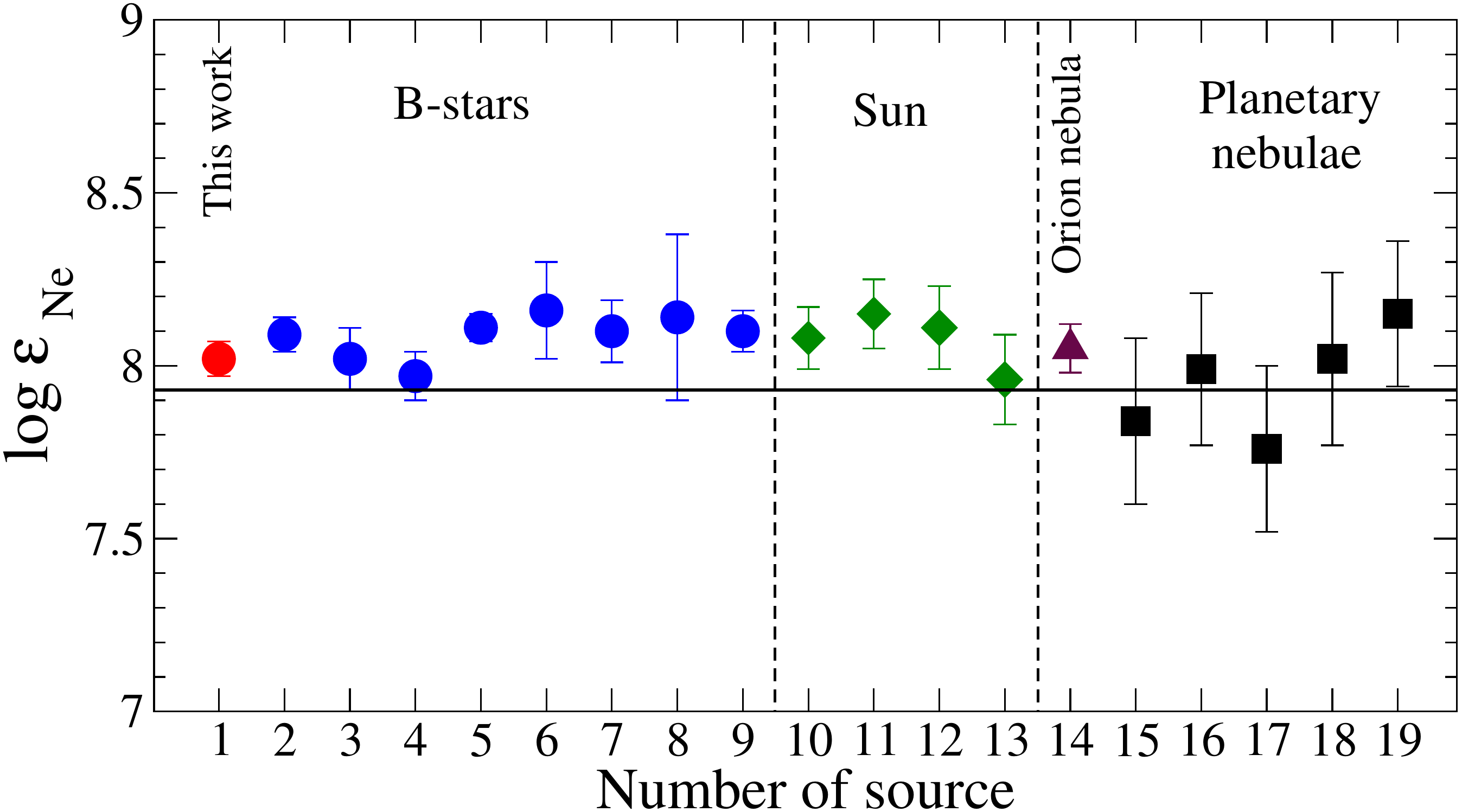}
 \caption{ Neon abundances compiled from different sources. Neon abundances from B-stars: (1) this work; (2) \citet{2012AA...539A.143N}; (3) \citet{2010PASJ...62.1239T}; (4) \citet{2008A&A...487..307M};
 (5) \citet{2006ApJ...647L.143C}; (6) \citet{2003A&A...408.1065H}; (7) \citet{2000MNRAS.318.1264D}; (8) \citet{1999ApJ...519..303S}; (9) \citet{1994A&A...282..867K}.
 Transition region of the quiet Sun: (10) \citet{2018ApJ...855...15Y} with oxygen abundance adopted from \citet{2009ARA&A..47..481A} and (11) \citet{2018ApJ...855...15Y} with the oxygen abundance adopted from \citet{2011SoPh..268..255C}.  The neon abundance in the solar atmosphere derived from UV spectrum of a solar flare: (12) \citet{2007ApJ...659..743L}. Solar wind observations: (13) \citet{2007A&A...471..315B}.
 The neon abundance in the Orion nebula: (14) \citet{2004MNRAS.355..229E}.
 Galactic disc planetary nebulae: 
 (15) \citet{2007MNRAS.381..669W}; (16) \citet{2006A&A...457..189P}; (17) \citet{2006ApJ...651..898S}; (18) \citet{2003A&A...409..619M}; (19) \citet{2003MNRAS.345..186T}.
   The solar abundance (7.93$\pm$0.10) as recommended by \citet{2009ARA&A..47..481A} is indicated by a horizontal line. 
 }
 \label{Compilation}
 \end{center}              
 \end{figure*}

  We compiled several neon abundance results recently obtained for B-stars, the Sun and the planetary nebulae (PNe) in the solar vicinity (Figure~\ref{Compilation}).
  
  Our mean neon abundance from B-stars, log~$\epsilon_{\rm Ne}$ = 8.02$\pm$0.05, is in a good agreement with the solar photospheric neon abundance (8.08$\pm$0.09), calculated from 
  recently revised ratio of neon to oxygen (Ne/O = 0.24$\pm$0.05) in the transition region of the quiet Sun \citep{2018ApJ...855...15Y}, where 
  the oxygen abundance, log~$\epsilon_{\rm O}$ = 8.69$\pm$0.05~dex, from \citet{2009ARA&A..47..481A} was adopted. 
  The average quiet Sun does not show any FIP effect \citep{2005A&A...439..361Y}, so these measurements are the most representative for the Sun. On the other hand, these measurements depend on the adopted oxygen solar abundance. For comparison, we took Ne abundance, log~$\epsilon_{\rm Ne}$ = 8.11$\pm$0.12~dex, derived from UV spectrum of a solar flare \citep{2007ApJ...659..743L} as the absolute abundance of Ne in the solar atmosphere. Our result is by 0.09~dex lower than that, although still consistent within the error bars. 
  Figure~\ref{Compilation} also presents the comparison with  
  the solar neon abundance from solar wind observations, log~$\epsilon_{\rm Ne}$ = 7.96$\pm$0.13~dex, \citep{2007A&A...471..315B}, that is by 0.06~dex lower that our result.
  
 The H\ii\ regions are relatively young objects. The nearest Galactic H\ii\ region is the Orion nebula. 
  Its chemical composition has been traditionally considered as the standard reference for the ionized gas in the solar neighborhood.
 \citet{2004MNRAS.355..229E} derived abundance in the Orion nebula gas for a large number of ions of different elements, including neon. 
They noticed that the abundances obtained from recombination lines
are larger than those derived from collisionally excited lines.
For example, the difference can reach $\sim$0.3~dex for neon assuming the temperature fluctuations ($t^2$ = 0.032$\pm$0.014).
We adopted the value of neon abundance in the Orion nebula gas
8.05$\pm$0.07~dex obtained from collisionally excited lines and assuming the temperature fluctuations ($t^2$ = 0.022$\pm$0.002), as recommended by \citet{2004MNRAS.355..229E}. 
Their value is only by 0.03~dex higher than the neon abundance in our B-stars (Fig.~\ref{Compilation}).

   Planetary nebulae (PNe) are expanding shells of the luminous gas expelled by dying stars of low and intermediate masses (LIMS). 
 They stem from
objects that have lifetimes up to Gyrs. The ionized gas surrounding the central star shows emission lines of highly ionized species from which the abundances can be derived. 
  The neon abundance can be compared and related to the results from stellar abundance analysis, since neon (and oxygen) originates from 
   primary nucleosynthesis in massive stars ($\geq$10~M$_\odot$), and is therefore nearly independent on the evolution of LIMS, the progenitor stars of PNe \citep{1989MNRAS.241..453H, 2004AJ....127.2284H}. 
    For comparison, we collected several studies, where the neon abundances were obtained in PNe.  We would not say that the abundances from PNe are very accurate, because the differences between abundances from collisionally excited lines and optical recombination lines can be much higher in PNe compared to those of H\ii\ regions.
For example, according to \citet{2007MNRAS.381..669W}, in four Galactic disc PNe (He~2-118, H 1-35, NGC~6567, and M~1-61) the mean neon abundance calculated from collisionally excited lines is 7.78$\pm$0.23, while it is
8.72$\pm$0.60 from optical recombination lines.   
 Figure~\ref{Compilation} presents neon abundances from different sources, including five studies focused on PNe.  
In those studies, where it was mentioned, we adopted neon abundances obtained from collisionally excited lines, e.g. \citet{2003MNRAS.345..186T, 2007MNRAS.381..669W}. The mean value of log~$\epsilon_{\rm Ne}$ = 7.84$\pm$0.24 was calculated on the basis of neon abundances in six Galactic disc PNe \citet{2007MNRAS.381..669W}.    
The mean value of the neon abundance log~$\epsilon_{\rm Ne}$ = 7.99$\pm$0.22 dex was calculated from six PNe (Hu~1-2, IC~418, NGC~40, NGC~2440, NGC~6543, NGC~7662) with Galactocentric distances from 7.9 kpc to 8.9 kpc from summary of \citet{2006A&A...457..189P}. 
The value of log~$\epsilon_{\rm Ne}$ = 7.76$\pm$0.24~dex was derived from 16 Galactic planetary nebulae located at the Galactocentric distances from 8.0 to 8.9 kpc in \citet{2006ApJ...651..898S}. 
The log~$\epsilon_{\rm Ne}$ = 8.02$\pm$0.25~dex was obtained from \citet{2003A&A...409..619M} from three PNe (NGC 6543, NGC 7027, NGC 7662) which are located at the distances no more than 1.0~kpc.  
We adopted the mean value log~$\epsilon_{\rm Ne}$ = 8.15$\pm$0.21~dex, which was calculated from 16 Galactic PNe from \citet{2003MNRAS.345..186T}. 

  Our neon abundance from B-stars is consistent within the error bars with the all above-mentioned measurements of neon in PNe. However, the scatter is quite large for neon abundance obtained from PNe.

\section{Conclusions}\label{Sect:Conclusions}

  We present a neon model atom based on improved atomic data, the NLTE analysis for Ne\ione\ and Ne\ii\ lines in the models representing the atmospheres of B-stars, and the consecutive determination of neon abundances in a sample of twenty-four B-stars in the solar neighborhood. 
The motivation of this work is to contribute to the solution of the 'Solar Model Problem'.
Of course, only solar indicators can ever be a probe for the solar photospheric neon abundance.
However, the counting all the physical effects makes the determination of solar neon abundance directly from solar indicators complicated. We believe that the indirect indicators, for example, B-stars from solar neighborhood could provide some indirect constraints on solar neon abundance. 
 We obtained the mean neon abundance from twenty-four B-stars in solar neighborhood is log~$\epsilon_{\rm Ne}$ =  8.02$\pm$0.05. According to this result and considering Galactochemical evolution and radial stellar migration, it seems unlikely that the solar neon abundance is 8.29 \citep{2005ApJ...631.1281B} that would be required to diminish the 'Solar Model Problem' in helioseismology.

  The model atom for Ne\ione\ -- Ne\ii\  has been constructed by using the most up-to-date atomic data and the non-local thermodynamic equilibrium (NLTE) line formation analysis was performed in classical 1D atmospheric models of B-type stars with the code \textsc{detail}. The use of the experimental oscillator strengths recently measured by Piracha et al. (2015) leads to smaller line-by-line scatter for the most investigated stars. 
   
   Our modeling predicts strong deviations from LTE for all considered Ne\ione\ lines and negative abundance corrections in the atmospheres with 10\,400 $\leq$ \Teff $\leq$ 30\,400~K.
   For example, for Ne\ione\ 6402~\AA\ line in the atmosphere with the parameters 17\,500 / 3.8, corresponding to $\iota$~Her, the difference between NLTE and LTE abundances, $\Delta_{\rm NLTE}$, can reach up to $-$1.1~dex. 
   We find that the main feature of Ne\ione\ is the extremely small photoionization cross-sections of the 3s levels, that leads to appreciable overpopulation of these levels.
   In contrast, the deviations from LTE for the most Ne\ii\ lines are small and do not exceed 0.11~dex in the absolute value. 
  
   We perform chemical abundance analyses of 24 well studied B-type stars in the solar vicinity in the temperature range 10\,400 $\leq$ \Teff $\leq$ 33\,400~K, using the the most recent  data of experimental transition probabilities taken from \citet{2015CaJPh..93...80P}.  Our study is based on high-resolution and high S/N ratio stellar spectra, where S/N ratio is higher than 1000 for some of them. The abundances of neon are determined from Ne\ione\ and Ne\ii\ lines in the optical and near-IR with LTE and NLTE line formation scenarios. 
   We find that, for each star, NLTE leads to smaller line-to-line scatter. For five stars with both Ne\ione\ and Ne\ii\ lines observed, NLTE provides ionization balance within 0.05~dex, while the LTE abundance difference can be as large as 0.50~dex in absolute value (e.g. HR~2928).
   
   We find that the Ne\ione\ lines are quite sensitive to electron-collisional data variations. 
   The using of R-matrix electron impact excitation calculations leads to strengthening of all Ne\ione\ lines by about $\sim$0.15~dex compared to 
   the case, where the approximation formula by \citet{1962ApJ...136..906V} is applied for allowed transitions and the effective collision strength $\Omega_{ij}$ = 1 for the forbidden ones.
   
   Our result is in accordance with previous neon estimations in B-stars \citep{1994A&A...282..867K, 1999ApJ...519..303S, 2000MNRAS.318.1264D, 2003A&A...408.1065H, 2006ApJ...647L.143C, 2008A&A...487..307M, 2010PASJ...62.1239T, 2012AA...539A.143N}.
   The mean neon abundance from B-stars is also in agreement with the solar photospheric neon abundance, log~$\epsilon_{\rm Ne}$ = 8.08$\pm$0.09, calculated from
   the transition region of the quiet Sun \citep{2018ApJ...855...15Y}, given that the oxygen abundance, log~$\epsilon_{\rm O}$ = 8.69$\pm$0.05~dex \citep{2009ARA&A..47..481A} is adopted.

\software{DETAIL \citep{detail}, SynthV\_NLTE \citep{2016MNRAS.456.1221R}, BINMAG \citep{2018ascl.soft05015K}}.

\acknowledgments
  
  This work was supported by the National Natural Science Foundation of China under grant Nos. 11988101, 11890694, U1631105 and
National Key R$\&$D Program of China No. 2019YFA0405502. 
 We thank the anonymous referee for valuable suggestions
and comments and Oleg Zatsarinny for providing us the
data on effective collision strengths for the Ne\ione\ transitions published in \citet{2009PhST..134a4020Z}.
  This research used the facilities of the Canadian Astronomy Data Centre operated by the National Research Council of Canada with the support of the Canadian Space Agency. 
  We made use of the NORAD Atomic Data, NIST, SIMBAD, VALD, and UK APAP Network databases.

\bibliography{Neon}
\bibliographystyle{aasjournal}

\end{document}